\ifx\destination\undefined
  \def\destination{aanda}  
\fi

\def\aanda{aanda}

\RequirePackage{snapshot}
\PassOptionsToPackage{dvipsnames}{xcolor}
\PassOptionsToPackage{colorlinks=True,allcolors=RoyalBlue}{hyperlink}

\ifx\destination\aanda
  \documentclass[ut8,T1,longauth]{aa}
  \usepackage{natbib}
  \bibpunct{(}{)}{;}{a}{}{,} 
\fi

\usepackage[utf8]{inputenc}
\usepackage[T1]{fontenc}

\usepackage{siunitx}
\DeclareSIPrefix\micro{\text{\textmu}}{-3}
\usepackage{graphicx}
\usepackage[title]{appendix}
\usepackage{booktabs}
\usepackage[skip=4pt]{caption}
\usepackage{here}
\graphicspath{{gfx/}}

\usepackage{lipsum}
\usepackage{pgf}
\usepackage{import}
\usepackage{tikz}

\usepackage{url}
\usepackage{xspace}
\usepackage[colorlinks=True,
            linkcolor=violet,         
            citecolor=RoyalBlue,    
            filecolor=violet,         
            urlcolor=violet           
           ]{hyperref}
\usepackage[nameinlink,capitalize]{cleveref}
\crefname{section}{Sect.}{Sects.} 
\usepackage{aas_macros} 

\usepackage{textcomp}
\usepackage{ulem}[normalem]

\usepackage[switch]{lineno}

\newcommand{\KM}[1]{\textcolor{magenta}{[#1]}}

\newcommand{\numb}[1]{\textcolor{orange}{#1}}
\renewcommand{\numb}[1]{\textcolor{black}{#1}}
\renewcommand{\KM}[1]{\textcolor{black}{#1}}
\newcommand{\genoid}{\texttt{Genoid}\xspace}
\newcommand{\eproc}{\texttt{Eproc}\xspace}
\newcommand{\miriade}{\texttt{Miriade}\xspace}
\newcommand{\ssodnet}{\texttt{SsODNet}\xspace}

\newcommand{\topcat}{\texttt{TOPCAT}\xspace}

\newcommand{\parallelsum}{\mathbin{\!/\mkern-5mu/\!}}
\usepackage[acronym]{glossaries}
\newacronym{sso}{SSO}{Solar System Object}
\newacronym{vwba}{VWBA}{Very Wide Binary Asteroid}
\newacronym{HST}{HST}{Hubble Space Telescope}
 
\makeglossaries

\begin{document}

\ifx\destination\aanda
  \title{Orbits of very distant asteroid satellites\thanks{Based on observations made with ESO Telescopes at the La Silla Paranal Observatory under programme ID 074.C-0052, 71.C-0669, 072.C-0753}}
  \subtitle{}
  \titlerunning{Orbits of very wide binary asteroids}
  \authorrunning{Minker et al.}

  \author{
    K.~Minker\inst{\ref{i:oca}}    \and 
    B.~Carry\inst{\ref{i:oca}}     \and 
    F.~Vachier\inst{\ref{i:imcce}} \and 
    P.~Scheirich\inst{\ref{i:Ondrejov}}  \and 
    P.~Pravec\inst{\ref{i:Ondrejov}}  \and 
    T.~M\"{u}ller\inst{\ref{i:maxplanck}}  \and 
    A.~Mo\'or\inst{\ref{i:Konkoly}}  \and 
    C.~Arcidiacono\inst{\ref{i:inaf}} \and 
    A.~Conrad\inst{\ref{i:lbto}}   \and 
    C.~Veillet\inst{\ref{i:lbto}}  \and 
    S.~A.~Jacobson\inst{\ref{i:michigan}}  \and 
    M.~Marsset\inst{\ref{i:ESO}}   \and 
    W.~J.~Merline\inst{\ref{i:swri}}  \and 
    P.~Tamblyn\inst{\ref{i:swri}}  \and 
    M.~E.~Brown\inst{\ref{i:caltech}}  \and 
    D.~Pray\inst{\ref{i:sugarloag}}  \and 
    R.~Montaigut\inst{\ref{i:opera}}  \and 
    A.~Leroy\inst{\ref{i:opera}}  \and 
    C.~Gillier\inst{\ref{i:chamonix}}  \and 
    P. Ku\v{s}nir\'ak\inst{\ref{i:Ondrejov}}  \and 
    K.~Hornoch\inst{\ref{i:Ondrejov}}  \and 
    M.~Hus\'arik\inst{\ref{i:skalnate}}  \and 
    V.~Benishek\inst{\ref{i:belgrade}}  \and 
    W.~Cooney\inst{\ref{i:sonoita}}  \and 
    J.~Gross\inst{\ref{i:sonoita}}\thanks{Deceased}  \and 
    D.~Terrell\inst{\ref{i:sonoita}, \ref{i:swri}}  \and 
    E.~Jehin\inst{\ref{i:trappist}}  \and 
    J.~Vil\'agi\inst{\ref{i:modra}}  \and 
    \v{S}. Gajdo\v{s} \inst{\ref{i:modrastefan}}  \and 
    V.~Chiorny\inst{\ref{i:kharkiv}}  \and 
    B.~Christmann\inst{\ref{i:soucieu}}  \and 
    J.~Brinsfield\inst{\ref{i:via capote}}  \and 
    C.~Dumas\inst{\ref{i:TMT}}  \and 
    B.~L.~Enke\inst{\ref{i:swri}}  \and 
    D.~D.~Durda\inst{\ref{i:swri}}  \and 
    J.~C.~Christou\inst{\ref{i:lbto}, \ref{i:noirlab}} \and 
    W.~M.~Grundy\inst{\ref{i:lowell}} \and
    L.~M.~Close\inst{\ref{i:steward}} \and
    S.~B.~Porter\inst{\ref{i:swri}} 
    }


  \institute{
    Universit{\'e} C{\^o}te d'Azur, Observatoire de la C{\^o}te d'Azur, CNRS, Laboratoire Lagrange, Bd de l'Observatoire, CS 34229, 06304 Nice cedex 4, France
    \email{kate.minker@oca.eu}
    \label{i:oca}
    \and 
    IMCCE, Observatoire de Paris, PSL Research University, CNRS, Sorbonne Universit{\'e}s, UPMC Univ Paris 06, Univ. Lille, 75014 Paris, France
    \label{i:imcce}
    \and
    Astronomical Institute AS CR, Fri\v{c}ova 298, Ond\v{r}ejov, Czech Republic
    \label{i:Ondrejov}
    \and
    Max-Planck-Institut f\"{u}r extraterrestrische
   Physik, Giessenbachstra{\ss}e, Postfach 1312, 85741 Garching, Germany
    \label{i:maxplanck}
    \and
    Konkoly Observatory, HUN-REN Research Centre for Astronomy and Earth Sciences, Konkoly-Thege Mikl\'os \'ut 15-17, H-1121 Budapest, Hungary
    \label{i:Konkoly}
    \and  
    INAF-Osservatorio Astronomico di Padova, Vicolo dell'Osservatorio 5, 35122 Padova Italy
    \label{i:inaf}
    \and  
    Large Binocular Telescope Observatory, University of Arizona, Tucson, AZ 85721, USA
    \label{i:lbto}
    \and 
    Department of Earth \& Environmental Sciences, Michigan State University, East Lansing, MI 48824, USA
    \label{i:michigan}
    \and  
    European Southern Observatory (ESO), Alonso de Cordova 3107, 1900, Casilla Vitacura, Santiago, Chile
    \label{i:ESO}
    \and
    Southwest Research Institute, 1301 Walnut St. \#400, Boulder, CO 80302, USA
    \label{i:swri}
    \and
    Division of Geological and Planetary Sciences, California Institute of Technology, Pasadena, CA 91125, USA
    \label{i:caltech}
    \and
    Sugarloaf Mountain Observatory 
    \label{i:sugarloag}
    \and
    OPERA Observatory
    \label{i:opera}
    \and
    Chamonix
    \label{i:chamonix}
    \and 
    Astronomical Institute of the Slovak Academy of Sciences, SK-05960 Tatransk\'{a} Lomnica, The Slovak Republic
    \label{i:skalnate}
    \and
    Belgrade Astronomical Observatory, Volgina 7, 11060 Belgrade 38, Serbia.
    \label{i:belgrade}
    \and
    Sonoita Research Observatory
    \label{i:sonoita}
    \and
    Space sciences, Technologies \& Astrophysics Research (STAR) Institute, University of Liege, Liege, Belgium
    \label{i:trappist}
    \and
    Modra Observatory, Department of Astronomy, Physics of the Earth, and Meteorology, FMPI UK, Mlynská dolina, Bratislava, 84248, Slovakia
    \label{i:modra}
    \and
    Department of Astronomy, Physics of the Earth, and Meteorology, FMPI, Comenius University, Mlynská Dolina F1, Bratislava, 84248, Slovakia
    \label{i:modrastefan}
    \and
    Institute of Astronomy of V.N. Karazin Kharkiv National University, Kharkiv 61022, Sumska Str. 35, Ukraine
    \label{i:kharkiv}
    \and
    Soucieu-en-Jarrest
    \label{i:soucieu}
    \and
    Via Capote Observatory, Thousand Oaks, CA 91320, USA
    \label{i:via capote}
    \and
    Royal Observatory Edinburgh, Blackford Hill, Edinburgh, EH9 3HJ, United Kingdom
    \label{i:TMT}
    \and
    Currently at NSF's NOIRLab/Gemini Observatory
    \label{i:noirlab}
    \and
    Lowell Observatory, 1400 W. Mars Hill Rd., Flagstaff AZ 86001
    \label{i:lowell}
    \and
    Steward Observatory, N420, Department of Astronomy, University of Arizona, 933 N. Cherry Ave. Tucson, AZ 85721
    \label{i:steward}
}

  \date{Received date / Accepted date }
  \keywords{%
     Methods: observational --
     Minor planets, asteroids: individual: (379) Huenna, (2577) Litva, (3548) Eurybates, (3749) Balam, (4674) Pauling, (22899) Alconrad, (17246) Christophedumas
     }
  \abstract
   {The very wide binary asteroid (VWBA) population is a small
   subset of the population of known binary and multiple asteroids
   made of  
   systems 
   with very widely orbiting satellites and long orbital periods, on the order of tens to hundreds of days. 
   The origin of these systems is debatable, and 
   most 
   members of this population are poorly characterized.}
   {We aim to develop orbital solutions for some members of the VWBA population, allowing us to constrain possible formation pathways for this unusual population.}
   {We have compiled all available high-angular resolution imaging
   archival data of VWBA systems from large ground- and space-based 
   telescopes. 
   We measure the astrometric positions of the satellite relative to the 
   primary at each epoch and analyze the dynamics of the satellites using the \genoid 
   genetic algorithm. Additionally, we use a NEATM thermal model to estimate the diameters of two systems, and we model the orbit of Litva's inner satellite using photometric lightcurve observations. }
   {We determine the effective diameters of binary systems (17246) Christophedumas and (22899) Alconrad to be $4.7\pm0.4$km and $5.2\pm0.3$ km respectively. We determine new orbital solutions for \KM{five} systems, (379) Huenna, (2577) Litva, (3548) Eurybates, (4674) Pauling, and (22899) Alconrad.
   We find a significantly eccentric ($e=0.30$)
   best-fit orbital solution for the outer satellite of (2577) Litva, moderately eccentric ($e=0.13$)
   solutions for (22899) Alconrad,
   and a nearly circular solution for (4674) Pauling ($e=0.04$).
   We also confirm previously reported orbital solutions for (379) Huenna and (3548) Eurybates.}
   {It is unlikely that BYORP expansion could be solely responsible for the
   formation of VWBAs, as only (4674) Pauling matches the necessary
   requirements for active BYORP expansion.
   It is possible that the satellites of these systems were formed
   through YORP spin-up and then later scattered onto very wide orbits.
   Additionally, we find that some members of the population are unlikely
   to have formed satellites through YORP spin-up,
   and a collisional formation history is favored.
   In particular, this applies to VWBAs within large dynamical families,
   such as (22899) Alconrad and (2577) Litva, or large VWBA systems such as (379) Huenna and NASA's Lucy mission target (3548) Eurybates.}
  \maketitle

\fi

\section{Introduction}%
\label{sec:introduction}%

There exists a small population of \glspl{vwba} -
binary or multiple systems with a satellite orbiting on a very distant orbit,
with typical orbital periods over 50 days, and semi-major axis typically $>10\%$ of the system's Hill radius ($r_H$).
Most members of this class were discovered by direct imaging,
through general surveys or targeted studies of asteroid families 
\citep{2002MerlineBalamDisco,
2003MerlineEsclangonaDisco,
2003MerlineAlconradDisco,
2004MerlinePaulingDiscoCBET,
2004TamblynChristophedumasDisco},
and several potential members of this class have been predicted through
lightcurve studies \citep{2019WarnerTrioOfVWBA}.
In 2012, (2577) Litva became the first member of these predicted \gls{vwba}
systems to be confirmed by direct imaging \citep{2013merlinelitvadiscoiauc, 2013merlinelitvadiscocbet}.
Analysis of asteroid observations in recent surveys 
\citep[with PanSTARRS and ESA's Gaia,][]{panstarrs, 2024arXiv240607195L}, and evidence in cratering records \citep{2024arXiv240518460H}
indicate that this population may be substantial,
but this has yet to be confirmed by observations. 

Wide binary systems have also been noted in the outer Solar system, categorized as ultra-wide Trans-Neptunian binaries (TNBs) \citep{parkerultrawide}. However, significant physical differences between the two populations (specifically, ultra-wide TNBs must have nearly equal-sized components, but no such systems have been observed in the inner Solar system) indicate that they are influenced by distinct formation and dynamical processes. Nevertheless, both populations are defined by observational criteria (a minimum separation of 0.5" at discovery for \cite{parkerultrawide}, or the somewhat looser condition that the satellite must be resolvable from the primary for VWBAs) which could falsely dichotomize what is actually a smooth distribution.

Although only a few members of the VWBA population have been identified, that does not necessarily indicate that these systems are rare,
as they are unusually difficult to observe.
As the possibility of detecting mutual eclipses by lightcurve indeed strongly decreases
with mutual separation \citep{vavilov2022},
the possibility of detecting mutual eclipses in VWBAs is nearly zero,
and these systems can typically only be detected by direct imaging under very favorable observing geometries \citep{pravec2012}. 

These systems should be easily detectable among near-Earth asteroids (NEAs) by radar, but such a population is yet to be identified. \cite{fangmargot2012} illustrate that planetary encounters affect wider binary systems more than narrow binary systems, considering cases with semi-major axis to primary-diameter ratios of $2<a/D_p<8$, with increased chances of instability for distant satellites. The VWBA population members are significantly more widely separated than this, with typical ratios of $a/D_P\approx50-100$. Extrapolating from the results of \cite{fangmargot2012}, any VWBAs that enter the NEA space would likely become unstable, and could be stripped of their satellites.

The origin of these systems is debated.
Early predictions suggested that these objects could be
escaping ejecta binaries (EEBs),
formed through the mutual capture of ejecta particles during the catastrophic
collision of a larger parent body \citep{2004Icar..167..382D}.
They could also form through YORP spin-up and get excited to their very wide orbits through various mechanisms
\citep{2011Icar..212..167P, 2014jacobsonVWBAs}.

Over the past decades, the Yarkovsky-O'Keefe-Radzievskii-Paddack
\citep[YORP, see, e.g.,][]{Vokrouhlicky2015YORP}
spin-up has become the favored mechanism to explain the formation of satellites
of small asteroids \citep{2008Natur.454..188W}.
Satellites formed through this mechanism 
are easily identified as this population
exhibits obvious signatures:
rapid primary spin periods $<4$\,h,
spin axis nearly perpendicular to the systems heliocentric orbital plane,
secondary-to-primary diameter ratio $D_s /D_p \approx 0.3$, a system semi-major axis at $\approx 1.5-2 D_p$  (although some may be slightly more distant),
and a preference for silicate compositions
\citep{2015-AsteroidsIV-Margot, 2023A&A...672A..48M}.
This mechanism is only effective for small ($<10$\,km) systems, and is expected to be responsible for the formation of most small binary asteroid systems.
While many systems show all of these characteristic signatures,
others present only a fraction of these qualities.
Most individual binary systems are not studied in detail,
and a rapidly-rotating primary for the system is often thought to be synonymous 
with YORP formation. This is likely an oversimplification, as an asteroid which 
has already acquired a satellite through another mechanism could still be 
susceptible to YORP spin-up of the primary.
Although the typical YORP binary system has been studied in detail 
\citep{2008Natur.454..188W, 2010Natur.466.1085P, 2011Icar..214..161J}
studies of population end members are typically less developed \citep{Agrusa2024DART}.

\begin{figure}[t]
\centering
\includegraphics[width=\columnwidth]{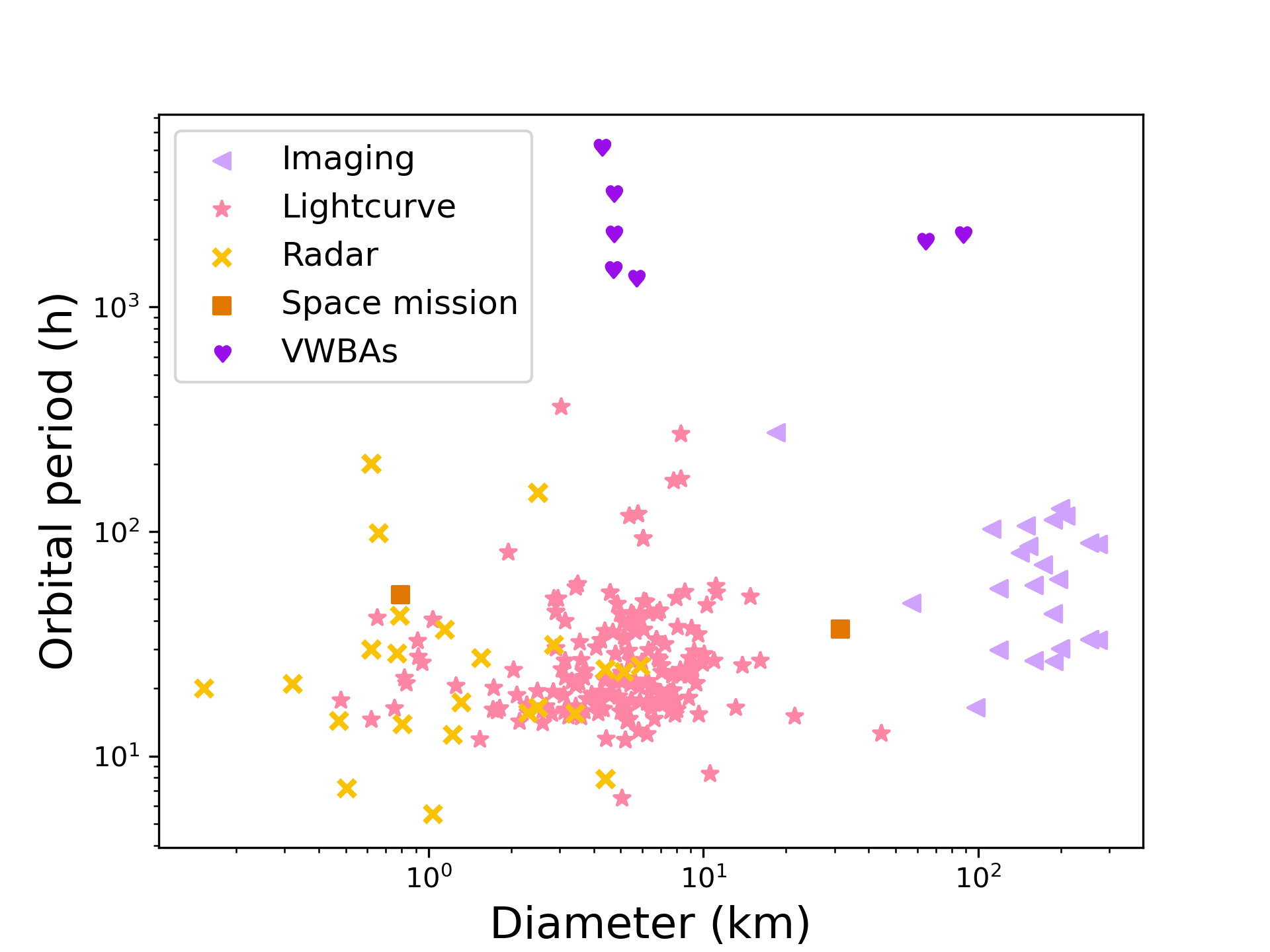}
\caption{Known binary asteroid systems, color- and symbol-coded by discovery method.
  Observational biases are clearly visible (for example, the populations of satellites discovered by imaging and lightcurves have no overlap),
  suggesting that gaps in the distribution (for example, in systems with $D_p\approx50$\,km or those with very long orbital periods) may simply be observational biases,
  not real attributes of the population. Systems discovered by stellar occultations are not shown, as the orbital periods are poorly constrained. Only systems with formal diameter estimates are included. As a result of this, systems with diameters $<1$\,km are underrepresented.
  }
\label{fig:binaryproperties}
\end{figure}

The recent flyby on November 1st, 2023 by NASA's Lucy mission \citep{lucygoodcitationlevison2021, 2024levisondinkineshnature}
of the main-belt asteroid (152830) Dinkinesh revealed the presence of
a previously undetected satellite, which was itself a contact binary.
Such a system had never been observed previously, and this serves as a reminder that the currently observed population may not be
representative of the true population of binary asteroids. 
Population end-members and unique systems, such as (5457) Queen's and (243) Ida,
are often 
discovered by unusual observational techniques
\citep[e.g., occultations, spacecraft flybys, see][]{cbet5318queens, chapman1995ida}. Both of these systems fall in the poorly populated intermediate size range of $D_p \approx 20-40$\,km
(\Cref{fig:binaryproperties}).
This suggests that there is more than meets the eye to the overall distribution of binary dynamics, and by extension the origins of these systems. 
Ill-studied portions of this dynamical space (for example, those with diameters between 10 and 100\,km) could be crucial to 
deepening our understanding of binary asteroid formation,
thereby expanding our view of ongoing dynamical processes affecting Solar system small bodies.

In this work, 
we present the first orbital solutions for 
\KM{three} members of this group, 
(2577) Litva,
(4674) Pauling,
and
(22899) Alconrad.
We also report \KM{updates and confirmations on existing orbital solutions for
(379) Huenna and (3548) Eurybates}. 
We summarize the properties of these targets in \Cref{tab:general}, 
some of which were extracted from \ssodnet\footnote{\url{https://ssp.imcce.fr/forms/ssocard}}
\citep{2023A&A...671A.151B}.

The manuscript is organized as following.
In \Cref{sec:Observations} we present the observational datasets used in this study,
alongside our data reduction techniques, and describe in 
\Cref{sec:Orbit} the orbit determination techniques and the parameters used in the orbital solutions.
We determine the physical properties of several of the targets in \Cref{sec:physical}.
The orbital solutions are presented in \Cref{sec:results}; and
in \Cref{sec:discussion} we discuss the implications of the presented
results on the formation history of \glspl{vwba}.


\begin{table*}[t]
\centering
  \caption{Basic parameters of known VWBA systems, from the literature. 
    \label{tab:general}
  }
  \begin{tabular}{llrrrrll}
  \hline\hline
    Number & Name & Tax. & $D_{\textrm{eff}}$ (km) & $\sigma_D$ & P (h) & Dyn. & Family\\
  \hline
    379 & Huenna & C$^1$ & 87.6$^2$ & 1.9 & 14.14$^{8}$ & Outer MB & Themis \\ 
    1509 & Esclangona & S$^1$ & 9.5$^2$ & 0.6 & 3.25285$^{9}$ & Hungaria & \\ 
    2577 & Litva & S$^1$ & 4.2$^2$ & 0.4 & 2.8129$^7$ & Mars-Crosser & Hungaria\\
    3548 & Eurybates & P$^1$ & 63.8$^3$ & 0.4 & 8.73$^{10}$ & Trojan & Eurybates\\ 
    3749 & Balam & S$^1$ & 4.7$^4$ & 0.2 & 2.804917$^{11}$ & Inner MB & Flora \\ 
    4674 & Pauling & S$^1$ & 4.7$^5$ & 0.1 & 2.5313$^{12}$ & Hungaria & \\ 
    17246 & Christophedumas & S$^6$ & 4.7$^7$ & 0.4 & $>10^{13}$ & Outer MB & Koronis \\
    22899 & Alconrad &  & 5.2$^7$ & 0.3 & 4.03$^{13}$ & Outer MB & Koronis \\
  \hline \\
  \end{tabular}\\
  \footnotesize{Listed are the asteroid name and number, taxonomic type (`Tax.'), effective diameter of the system (`$D_{\textrm{eff}}$', based off of an equal-volume sphere, except for Eurybates, where it is areal equivalent), uncertainty on the diameter ($\sigma_D$), rotation period ('P'), Dynamical class (`Dyn.'), and Family. The uncertainties on the diameter estimates are formal in most cases, and may be subject to additional systematic errors. References are as follows: 1: \cite{2022MaxTax}, 2: \cite{2023A&A...671A.151B} 3: \cite{2023PSJ.....4...18M}, 4: \cite{2011MasieroWISE}, 5: \cite{2022MyhrvoldWISE}, 
  6: \citet{2021A&A...652A..59S}, 
  7: This work, 8: \cite{2010MPBuWarnerHuenna}, 9: \cite{2009esclangonawarner}, 10: \cite{2021MPBu...48...13S}, 11: \cite{2020Icar..33613415P}, 12: \cite{2011MPBu...38...25W}, 13: \cite{polishook2011vwba}. }
\end{table*}

\begin{figure}[t]
\centering
\includegraphics[width=\columnwidth]{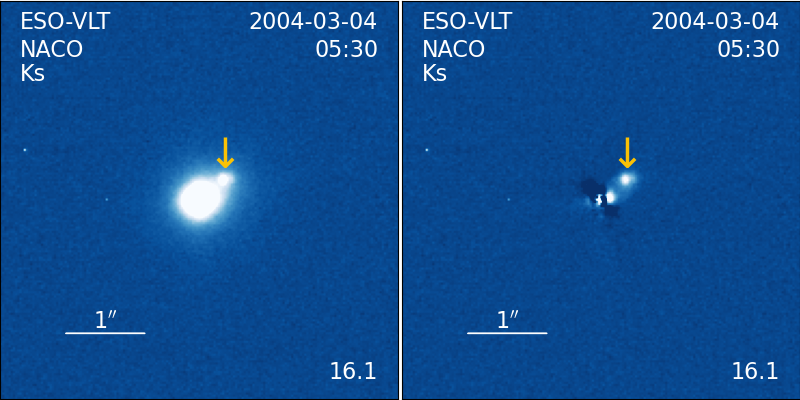}
\includegraphics[width=\columnwidth]{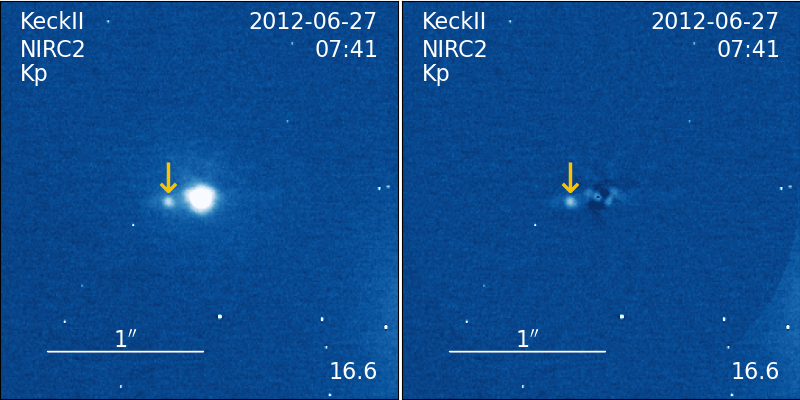}
\caption{Example of ground-based images for
  (4674) Pauling (top) and (2577) Litva (bottom).
  The two columns present the images before (left) and after (right) the application of halo-subtraction algorithm
  (see text).
  The arrows indicate the position of the satellite.
  The instrument and filter are indicated in the upper left corner of each image,
  the time of observation in the upper right corner, 
  and the apparent magnitude in the lower right corner.
  \label{fig:pauling}
  }
\end{figure}

\begin{figure}[t]
\centering
\includegraphics[width=0.49\columnwidth]{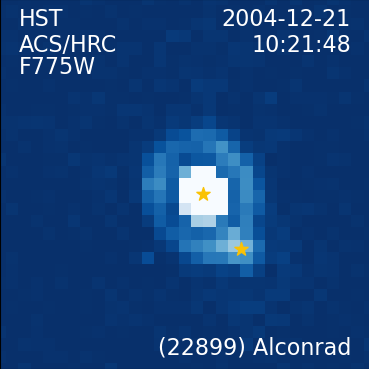}%
\includegraphics[width=0.49\columnwidth]{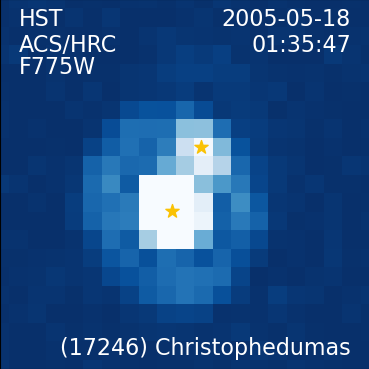}
\caption{Example of images of (22899) Alconrad (left) and (17246) Christophedumas (right)
  imaged by the \gls{HST}.
  The measured positions of the primary and satellite are marked with yellow stars. The satellite position overlaps an Airy ring from the primary PSF in both images.
  \label{fig:alconrad}
  }
\end{figure}

\section{Observational dataset\label{sec:Observations}}

We have identified eight \glspl{vwba} for which imaging data were available:
(379) Huenna, (1509) Esclangona, (2577) Litva, (3548) Eurybates, (3749) Balam,
(4674) Pauling, (17246) Christophedumas, and (22899) Alconrad. 
For both Huenna and Eurybates, orbital solutions were recently published by
\cite{2022Icar..38215013V} and \cite{2021-Brown-Eurybates, Nolleurybates2023}.
We present here a minor update to the orbital solution for Huenna
from that \cite{2022Icar..38215013V} containing an additional epoch of observations,
and a solution derived from the astrometry presented in \cite{2021-Brown-Eurybates} when discussing Eurybates. 

\subsection{Direct imaging of \gls{vwba} systems\label{ssec:images}}

To maximize the dataset, both in number of objects and observations
per system, we have searched the archives of all large
(8m-class) ground-based telescopes equipped with adaptive-optics cameras
as well as the \gls{HST},
for observations of \glspl{vwba}.
We found images of these systems in the archives of
the \gls{HST}, ESO VLT, Keck, and LBT observatories.
Many of these observations are from observing programs led by authors of this study.
Images taken at the ESO VLT used the NACO instrument
\citep{2003-SPIE-4841-Lenzen, 2003-SPIE-4839-Rousset}, 
those from Keck were acquired with the NIRC2 instrument
\citep{2004-AppOpt-43-vanDam}, 
observations at LBT used the PISCES instrument\footnote{\url{http://hdl.handle.net/20.500.12386/33745}}, and the \gls{HST} images were obtained with
the HRC channel of the ACS instrument \citep{2023acsi.book...23R}.
The observing programs of these observations can be found under the following PIDs.
From the European Southern Observatory: 074.C-0052 (PI: Marchis), 71.C-0669 (PI: Merline), 072.C-0753 (PI: Merline).
From the W. M. Keck Observatory: N086N2 (PI: Merline), K296N2L (PI: Armandroff), N255N2L (PI:Merline), A283N2L (PI: Merline), N131N2L (PI: Porter), C232 (PI: Brown), and several instances of observations conducted on engineering time. From the Hubble Space Telescope: 10165 and 9747 (PI: Merline). Some of these observations have been previously published with or without reports of satellite astrometry; see for example \cite{2002MerlineBalamDisco, 2003MerlineEsclangonaDisco, 2003MerlineAlconradDisco, 2004MerlinePaulingDiscoCBET, 2013merlinelitvadiscocbet, 2004TamblynChristophedumasDisco,  marchis2008eccentric}.

All ground-based images were processed using standard image calibration techniques, 
including dark subtraction, flat-fielding, identification of dead
and hot pixels, as well as halo-reduction algorithms
to maximize the detectability of the satellite
\citep{2008AA...478..235C, 2018Icar..309..134P}.
The astrometry of the satellite on each image was measured by fitting
a 2D Gaussian profile.
Examples of processed images can be found in \Cref{fig:pauling}.
In the case of (379) Huenna, most positions used here
are duplicated from \cite{2022Icar..38215013V}, which applied the same methodology. 

The processing and astrometry of images acquired with the \gls{HST} were performed using DOLPHOT
\citep{Dolphin2000, Dolphin2016}. 
Pre-computed ACS PSFs were used to fit the astrometric positions of the primary and satellite in each image, one high-quality image was used per epoch.
Only two systems were observed in this way: 
(17246) Christophedumas and (22899) Alconrad.
We had a positive detection of the satellite in all but one epoch,
resulting in the four and five positive detections for Christophedumas and Alconrad, respectively.
Examples images can be found in \Cref{fig:alconrad}, and a detailed description of the observational epochs, images used, and astrometric positions reported can be found in appendix \ref{app:obs:sat}.

Unfortunately, we found 
(1509) Esclangona and (3749) Balam to have insufficient datasets
for meaningful orbit determination.
In the case of Esclangona, the satellite was resolved on
only three dates, spread over nine years. This very low temporal-density dataset is not suitable for orbit determination.
In the case of Balam, the present dataset is significantly reduced from previous studies
\citep{marchis2008eccentric, vachier2012genoide}.
A known field-orientation issue with
Gemini/Hokupa'a\footnote{\url{https://www.gemini.edu/sciops/instruments/uhaos/uhaosIndex.html}}
prohibits the use of the discovery data of Balam's satellite
\citep{2002MerlineBalamDisco}.
Although the satellite is clearly resolved in these images,
the absence of a reliable field-orientation makes them unsuitable for orbital modeling.
The dataset for Balam is further reduced as we were not able to find the
satellite at the July 2003 and November 2004 epochs, reported in previous studies
\citep{marchis2008eccentric, vachier2012genoide}.
Although we are unsure of the specific image processing techniques used to obtain the position of the satellite in these images, the conspicuous lack of reported satellite magnitudes on these dates combined with the fact that the reported separations on these dates are extremely small suggests that the reported positions were likely approximative. 
It is possible that the satellite was
very close to the primary during these epochs, making it
impossible to resolve the system. Since no precise satellite positions can be
measured from these images, we exclude them from our dataset. \KM{We also found the dataset for (17246) Christophedumas to be insufficient to produce a robust orbital solution, see section \ref{sec:cdorb} for further information.}
We report the relative astrometry and photometry of all satellites in
\Cref{app:obs:sat}.

\subsection{Optical lightcurves\label{sec:lc}}

Two objects in our sample of VWBAs, (2577) Litva and (3749) Balam are triple systems and possess a closely-orbiting satellite in addition to their very wide satellite. For (2577) Litva, we construct a model for the closely-orbiting satellite in addition to the very wide satellite in order to fully characterize the system. This was not replicated for Balam, as the orbit of the outer satellite is poorly constrained.
In this work, we utilized the high-quality photometric observations of Litva that were taken in its four apparitions of 2009, 2010, 2012 and 2018.
The observations taken in the first two of the four years were published in \cite{pravec2012}.
In 2012, it was observed from April 16 to June 19, from following observatories:
Skalnat\'e Pleso (with a 0.61-m telescope on 10 nights),
Ond\v{r}ejov (0.65-m, 7 nights),
Via Capote (0.36-m, 4 nights),
Sonoita Research Observatory (0.5-m, 4 nights),
La Silla-TRAPPIST \citep[0.6-m, 4 nights,][]{2011Msngr.145....2J},
Modra (0.60-m, 2 nights), and 
Kharkiv (0.70-m, 2 nights).
In 2018, it was observed from September 3 to 16, from following observatories:
Sugarloaf Mountain (0.50-m, 4 nights),
OPERA (0.20-m, 3 nights),
Soucieu-en-Jarrest (0.20-m, 3 nights), and
Sopot (0.35-m, 1 night).
The observations from the individual stations were processed using the standard techniques (bias, dark and flat field corrections) and
photometrically reduced with the aperture photometry method using specific software packages utilized at the individual observatories.
The observational and reduction techniques utilized at the individual observatories were described in \cite{pravec2012, 2014Icar..233...48P, 2016Icar..267..267P, 2019pravecpairs, 2019sf2a.conf..471M} and in \cite{2019MPBu...46...87B}.
Generally, they were high-quality data with the RMS residuals (in the binary lightcurve decomposition; see below) of 0.013 and 0.015 mag in 2012 and 2018,
respectively.  The data taken with the 0.65-m telescope at Ond\v{r}ejov on the 7 nights in 2012 were absolutely calibrated in the Cousins R system with the \cite{1992AJ....104..340L} standards with the absolute calibration errors of 0.01 mag.  
The data from the other observatories were taken as relative or they were
calibrated in different photometric systems with a lower internal consistency and we used them as relative, taking the zero points of the magnitude scales on the individual nights as free parameters and adjusting them for the best fit in the lightcurve decompositions below.

\begin{figure}[t]
\centering
\includegraphics[width=\columnwidth]{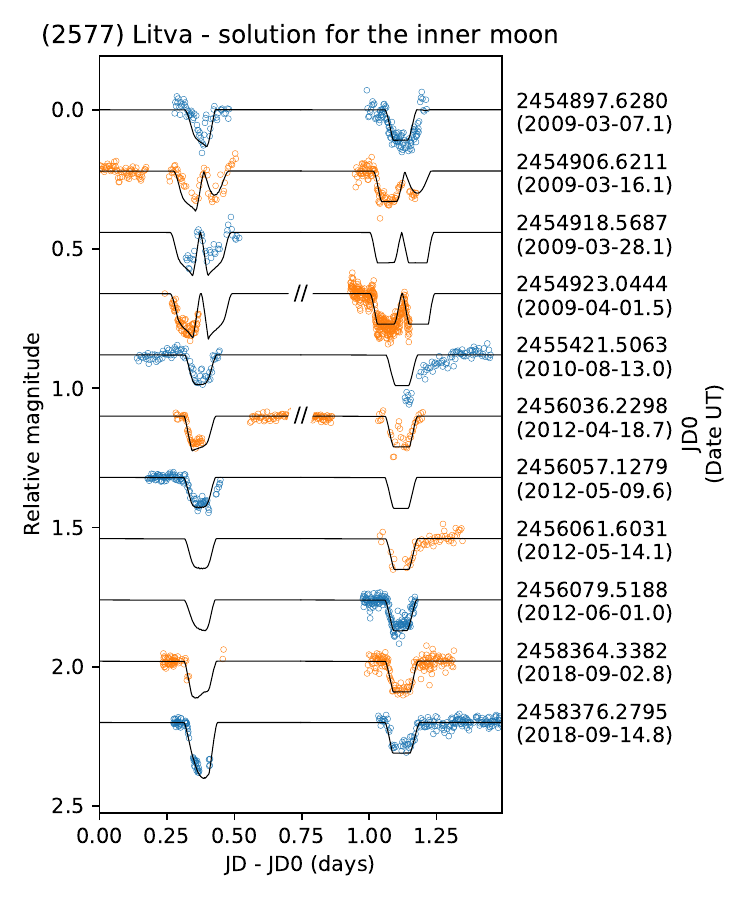}
\caption{%
Example of orbital (mutual event) lightcurves of the inner satellite of (2577) Litva observed in 2009, 2010, 2012 and 2018.  The observed data with the rotational lightcurves of the primary and the
outer satellite subtracted using the binary asteroid lightcurve decomposition method \citep[][and reference therein]{pravec2022}
are marked as points.
  The solid curve represents the synthetic lightcurve for the nominal solution
  for the inner moon.
  The primary and secondary events 
  are always shown on the left and right sides of the plots,
  respectively.
  In two cases where
  the secondary event was observed before the primary event, 
  we present them in reverse order, separated
  by the $\parallelsum$ symbol and one orbital period (1.4946 day).
  \label{fig:litvalightcurves}
}
\end{figure}

\subsection{Mid-infrared photometry\label{ssec:ir}}

Two of the systems here, (17246) Christophedumas and (22899) Alconrad,
had no published diameter estimate.
We thus searched for mid-infrared data to constrain their
diameter and albedo (discussed in \Cref{sec:physical}).
We found that both systems were observed during the course of the
NASA WISE survey \citep{wright2010}. They were also the
target of a NASA Spitzer program 
[P.~Tamblyn, PID:20538] with the 
IRAC instrument \citep{fazio2004}.

The Spitzer
observations were carried out using all four IRAC bands at 3.6, 4.5, 5.8, and 8.0{$\mu$m}. 
Christophedumas was measured on 2005 August 22, while the observation of Alconrad was made on 2005 December 27. 
To extract the flux densities of the asteroids, we used aperture photometry on the individual corrected basic calibrated data (CBCD) images produced by the IRAC pipeline and downloaded from the Spitzer Heritage Archive (AORKEY: 14818304 and 14818560 for Christophedumas and Alconrad, respectively).
The aperture radius was set to 2\,pixels 
($\sim$2.4\arcsec), 
and the sky level was estimated by computing a robust average in a radial annulus from 2 to 6\,pixels. We applied aperture correction using the appropriate factors from the IRAC Instrument Handbook v4.0\footnote{\url{https://irsa.ipac.caltech.edu/data/SPITZER/docs/irac/iracinstrumenthandbook/}} and pixel phase correction employing the \texttt{irac\_aphot\_corr} task\footnote{\url{https://irsa.ipac.caltech.edu/data/SPITZER/docs/dataanalysistools/tools/contributed/irac/iracaphotcorr/}} developed by the Spitzer Science Center. 
To derive the final photometry and the uncertainties, we computed the mean and the error of the mean of the individual flux density measurements using an outlier resistant method. All these data are summarized in \cref{app:obs:ir}, where the total uncertainties -- the quadratic sums of the measurement errors and the calibration uncertainties (IRAC Instrument Handbook) -- are also given.
For both systems, model calculations show that the IRAC1 and IRAC2 bands contain
significant contributions of reflected sunlight,
so we focused on the IRAC3 and IRAC4 bands.
The in-band fluxes were color-corrected
to obtain monochromatic flux densities. 
Ultimately, diameters of $4.7\pm0.4$\,km and $5.2\pm0.3$\,km were determined using the process described in section \ref{sec:diameters} for Christophedumas and Alconrad, respectively. A description of the diameter estimates for all systems considered in this study can be found in Table \ref{tab:general}.


\section{Orbit Determination\label{sec:Orbit}}

We determine the best-fit orbital solutions for the binary systems with the \genoid
algorithm \citep{2012AA...543A..68V}.
\genoid is a genetic algorithm that 
searches for the dynamical parameters of the orbital motion in a binary or multiple system through successive generations of randomly matched solutions. The first generation of solutions is drawn randomly over a large range of values for each parameter (mass of the components,
semi-major axis, etc). An ephemeris for each observing date is then computed using 
\eproc \citep{1998-IMCCE-Berthier}, the computation library behind
IMCCE online ephemerides \citep{2006-ASPC-351-Berthier,
2016MNRAS.458.3394B, 2018P&SS..164...79C}.
A fitness function based on $\chi^2$ 
is minimized during the process to determine the solution that
shows the lowest residuals between the observed and computed positions (hereafter “o-c”).
The high number of trial solutions 
(from 300,000-2,000,000 depending on the complexity of the system)
combined with the numerous successive generations (1000)
ensures both
a wide exploration of the parameter space and convergence toward the global
minimum of residuals.
A complete description of the method can be found in \citet{2012AA...543A..68V},
and other applications of this algorithm can be found in, e.g.,
\citet{2014Icar..239..118B},
\citet{2018Icar..309..134P},
\citet{2019A&A...623A.132C, 2021A&A...650A.129C},
\citet{2020A&A...641A..80Y}, and
\citet{2022Icar..38215013V}.
Uncertainties on variable parameters are statistical, not formal, and represent the range of values present in solutions where the RMS residuals are below the average 1 $\sigma$ threshold. Since most orbits calculated with \genoid have RMS residuals significantly below this threshold, this may result in the overestimation of these statistical uncertainties.

Due to the limited size of our datasets, for most systems we restrict the solutions
to simple Keplerian orbits, and all higher order terms
(external perturbation of the Sun and planets,
influences due to the non-spherical nature of the systems' primaries)
have been neglected. In other terms, the dynamical system reduces
to a restricted two-body problem. 
However, we wish to note that the very wide nature of these orbits
implies that the Sun and planets likely have a substantial influence
on the long-term orbital dynamics, and it will likely be necessary to fit
long-term orbital solutions incorporating future observations.
In the case of (379) Huenna and (3548) Eurybates, the dataset is much larger,
so we have included these perturbations,
as was previously described by \citet{2022Icar..38215013V}.
The mass of the satellite is also non-negligible,
such that the center of mass of the system is likely outside of the primary.
However, the one-pixel uncertainties on the position of the primary encompass
the anticipated location of the center of mass
(assuming equal density and albedo of all components),
and as such the center of mass can be approximated as the photocenter
of the primary for the present limited dataset. The masses reported in Section \ref{sec:results} represent the total system mass, assuming a Keplerian orbit. Similarly, in the case of the triple systems, the inner satellite may cause a discrepancy between the photocenters' and the systems' center of mass, but once again we estimate this discrepancy to be negligible, and therefore approximate the inner two components of the system as one central object.
The results of these orbital fits can be found in Section \ref{sec:results}.

\section{Physical models \label{sec:physical}}

\subsection{Diameter estimates \label{sec:diameters}}

One object from our sample, (17246) Christophedumas, does not have a diameter estimate reported in the literature, and that of (22899) Alconrad has only one reported estimate \citep{2011MasieroWISE} which does not include all available data. We have derived diameter and albedo estimates for these two asteroids as described in the following, using thermal observations described in Section \ref{ssec:ir}.

\subsubsection{(17246) Christophedumas \label{sec:CDdiam}}

We applied a simple thermal model for near-Earth asteroids
\citep[NEATM,][]{1998-Icarus-131-Harris}
to the derived fluxes presented in Table \ref{tab:spitzerchris}.
The NEATM analysis of the IRAC data returns 
a beaming parameter
of 1.165, a diameter of $4.8\pm0.1$\,km and a geometric V-band albedo of
$p_V=0.21\pm0.04$ when assuming
$H_V=13.93\pm0.2$. 
For the $H_V$ the ATLAS orange-band $H_o$ from \citet{2021Icar..35414094M} was used,
and translated into $H_V$ by the method of 
\citet[][\& private communication]{2022A&A...666A.190S}. See also
footnote 9 in \citet{2023A&A...670A..53M}. 

This diameter estimate represents
the effective diameter of the entire binary systems ($D_{\textrm{eff}}$) and uncertainties are formal, although these estimates maybe be subject to additional systematic uncertainties.
The best NEATM solution for the WISE data is found for a beaming parameter of 1.22, a size of $5.01\pm0.5$\,km and a geometric V-band albedo of
$0.19 
\pm0.04$.
The IRAC and WISE solutions are perfectly compatible, in beaming
parameter and
size-albedo solution. However, the IRAC-derived size and albedo has a higher accuracy. 
Assuming the 0.15 mag lower limit on the lightcurve amplitude
\citep{2011Icar..212..167P}, 
it could also be that the measurements by IRAC in 2005 and WISE in 2010 have seen the system at different rotational phases and also under a different aspect angle. 
Only a high quality spin-shape solution for the system would make it possible to constrain a single size-albedo solution.
As the IRAC (phase angle $\alpha$ of -21\degr)
and WISE data ($\alpha$=+21 deg) are taken before and after opposition,
respectively,
a reliable spin-shape solution would also make it possible to constrain the object's
thermal properties. A first attempt using a thermophysical model
\citep[TPM,][]{1996A&A...310.1011L, 1996A&A...315..625L,
1997A&A...325.1226L, 1998A&A...332.1123L, 1998A&A...338..340M}
assuming
the 10h lower limit on the rotation period \citep{2011Icar..212..167P},
and a spin-pole perpendicular
to the ecliptic plane gave a radiometric size (for the combined WISE \& IRAC
data set) of $4.7 \pm 0.4$\,km (of an equal-volume sphere), a geometric V-band
albedo $p_V$ of $0.22\pm0.04$ and a strong indication for a thermal inertia
in the range 4-50 J m$^{-2}$ s$^{-0.5}$ K$^{-1}$.
Considering a diameter ratio of 0.34 between the primary and the secondary,
this becomes diameters of 4.4\,km and 1.5\,km for the two components of the system. \KM{An illustration of the $\chi^2$ curves from the TPM fit can be found in Figure \ref{fig:CD_NEATM}}.

\subsubsection{(22899) Alconrad}

The 
NEATM 
analysis gave a beaming parameter
of 1.15, a size of $4.9 \pm 0.2$\,km and a geometric V-band albedo of
$0.26\pm0.04$
\citep[assuming $H_V=13.65\pm0.2$, derived following the same methods as described in Section \ref{sec:CDdiam} from $H_o=13.33$ from][]{2021Icar..35414094M}. 
 \cite{2011MasieroWISE} used 11 WISE-W3 and 7 W4 detections to derive
a NEATM-based size and albedo solution with D=5.68 ±0.47 km, $p_V=0.18 \pm0.03$
taking an H-mag of 13.70 and a fixed beaming parameter of $1.30 \pm0.16$.
 MPC archive lists additional detections, 13 in total (observatory code C51).

All W3 and W4 detections happened in the time period between 2010-Mar-10
12:23 and
2010-Mar-11 15:22 (mid-time: 2010-Mar-11 02:00, 2455266.58333:
r=2.91372 au, $\delta$=2.64959, $\alpha$= +19.8).
The W3 and W4 multi-band detections show some scatter in flux
but no W3-W4 synchronized variation as one would expect for
a shape-driven thermal lightcurve. Therefore, we calculated the
weighted average of the good-quality 
(flag `0A') W3 and the best-quality
W4 points (after applying color correction factors of 0.94 and 0.98
in W3 and W4, respectively) \citep{SzakatsColorCorrection}.
The results are: 6.09 $\pm$ 0.58 mJy at 11.1
micron,
and $16.96 \pm 1.14$ mJy at 22.64 micron.
A NEATM calculation leads to a size of $D=4.9 \pm 0.2$ km, geometric
V-band albedo of $p_V=0.25 \pm0.05$ (using the $H_V=13.65$, see above), and
a best-fit beaming parameter $\eta = 1.05$.
The IRAC and WISE solutions are compatible, in beaming parameter and
size-albedo solution. However, the IRAC-derived size and albedo has a
slightly higher accuracy. It should also be mentioned that the
radiometric solution is
related to the size of an equal-volume sphere for the combined binary
system.
As the lightcurve amplitude of the system is found to be between 0.14
and 0.19 mag
\citep{2011Icar..212..167P},
it could also be that the measurements by IRAC in 2005
and WISE in 2010 have seen the system at different rotational phases and also
under a different aspect angle. 
Only a high quality spin-shape solution for the system would make it possible to constrain a single size-albedo solution.
The IRAC
($\alpha$=+19.3\degr)
and WISE data ($\alpha$=+19.8\degr) are both taken after opposition, which makes
a study of thermal properties more challenging.

A first attempt using a TPM with a 4.03 h rotation period \citep{2011Icar..212..167P}, and a spin-pole perpendicular to the orbital plane of
the satellite $(\lambda,\beta)_{ECJ2000} = (330^o,-54^o)$, i.e., a retrograde sense
of rotation, gave a radiometric size (for the combined WISE \& IRAC
data set) of $5.2\pm0.3$, $p_V=0.24\pm0.03$ and an indication for a thermal inertia between 30 and 100 J m$^{-2}$ s$^{-0.5}$ K$^{-1}$.
Notably, the 4.03 h spin period may be unreliable, as an alternative estimation of 5.02 h is reported by \citep{2015AJ....150...75W}. This TPM
radiometric solution fits the data on an accepted level (reduced $\chi^2$
close to 1.0), but it is closely connected to the retrograde sense
of rotation of the primary derived from the satellite's orbit. 
When using a prograde sense of rotation with ($\lambda,\beta$) = $(330^o,+54^o)$ the
radiometric solution changes dramatically: the reduced $\chi^2$ (for the
best TPM solution in comparison with the combined WISE and IRAC data)
jumps to values above 4 which means that a prograde rotation
of the primary body is not compatible with the available
thermal measurements (see also $\chi^2$ plot in the appendix, \KM{Figure \ref{fig:AC_NEATM})}.
However, also the retrograde TPM solution should be taken with care
as it depends strongly on the assumed spin-pole and thermal properties.
In addition, the best-fit thermal inertia (which determines
the final radiometric size) might be affected by unknown shape effects.
Such radiometric studies suffer therefore from errors in the model
input parameters which are difficult to quantify.

\subsection{Inner satellite of (2577) Litva}

We constructed a model of the orbit of the inner moon of Litva using the technique of \citet{2009Icar..200..531S}, modiﬁed to allow for precession of the pericenter in the case when modeling the eccentric orbit (see below), which was further developed by \citet{2015Icar..245...56S} and \citet{2021Icar..36014321S}. 
We outline the basic points of the method below, but we refer the reader to the above references for details of the technique.

The binary asteroid components were represented with a sphere (for the inner moon) and oblate ellipsoid (for the primary), orbiting each other on a circular orbit.
We choose the circular orbit for simplicity, as the upper limit on the eccentricity is low (see below).
The motion was assumed to be Keplerian.
The spin axis of the primary was assumed to be normal to the mutual orbital plane of the components
(i.e., we assumed the mutual orbit is in the equatorial plane of the primary).
The shapes were approximated with 1016 and 252 triangular facets for the primary
and the secondary, respectively.  The components were assumed to have the same albedo and to be free of albedo features \citep[see][for discussion on why albedo features can be neglected]{2001Icar..153...24K}.
The brightness of the system as seen by the observer was computed as a sum of contributions from all visible facets using a ray-tracing code that checks which facets are occulted by or are in shadow from the other body.
A combination of Lommel-Seeliger and Lambert scattering
laws was used \citep[see, e.g.,][]{2002-AsteroidsIII-2.2-Kaasalainen}.

The photometric data were analyzed using the binary asteroid lightcurve decomposition method (\cite{pravec2022}, and references therein). There were present two rotational lightcurves in the data, one belonging to the primary and the other belonging to the outer satellite, with spin periods of 2.8129 and 5.6818 h, respectively \citep{2016Icar..267..267P}.  Both rotational lightcurves were subtracted from the photometry data
with the lightcurve decomposition technique and we obtained the orbital lightcurve component, showing mutual events (eclipses and/or occultations) between the primary and the inner satellite.

The solution for the parameters of the orbit of the inner satellite given in Table~\ref{tab:litvaorbit} was obtained by fitting our model to the orbital lightcurve component from the apparitions 2009, 2010, 2012 and 2018 simultaneously. To account for the presence of the third body, a total light ﬂux scattered towards the observer was computed as $I_1 + I_2 + I_3$, where $I_1$ is the light ﬂux from the $i$th body. We set $I_3/(I_1+I_2) = 0.053$ (i.e., the square of the size ratio between the outer satellite and the combined light of the primary and inner satellite, see \cref{sec:litvaorb}).
The upper limit on the eccentricity was estimated by fitting the orbital lightcurve component from the best-covered apparition (2012) only. Examples of the orbital lightcurve component of the observed data together with a synthetic lightcurve for the nominal solution are presented in \cref{fig:litvalightcurves}.

\begin{figure}[t]
\centering
\includegraphics[width=\columnwidth]{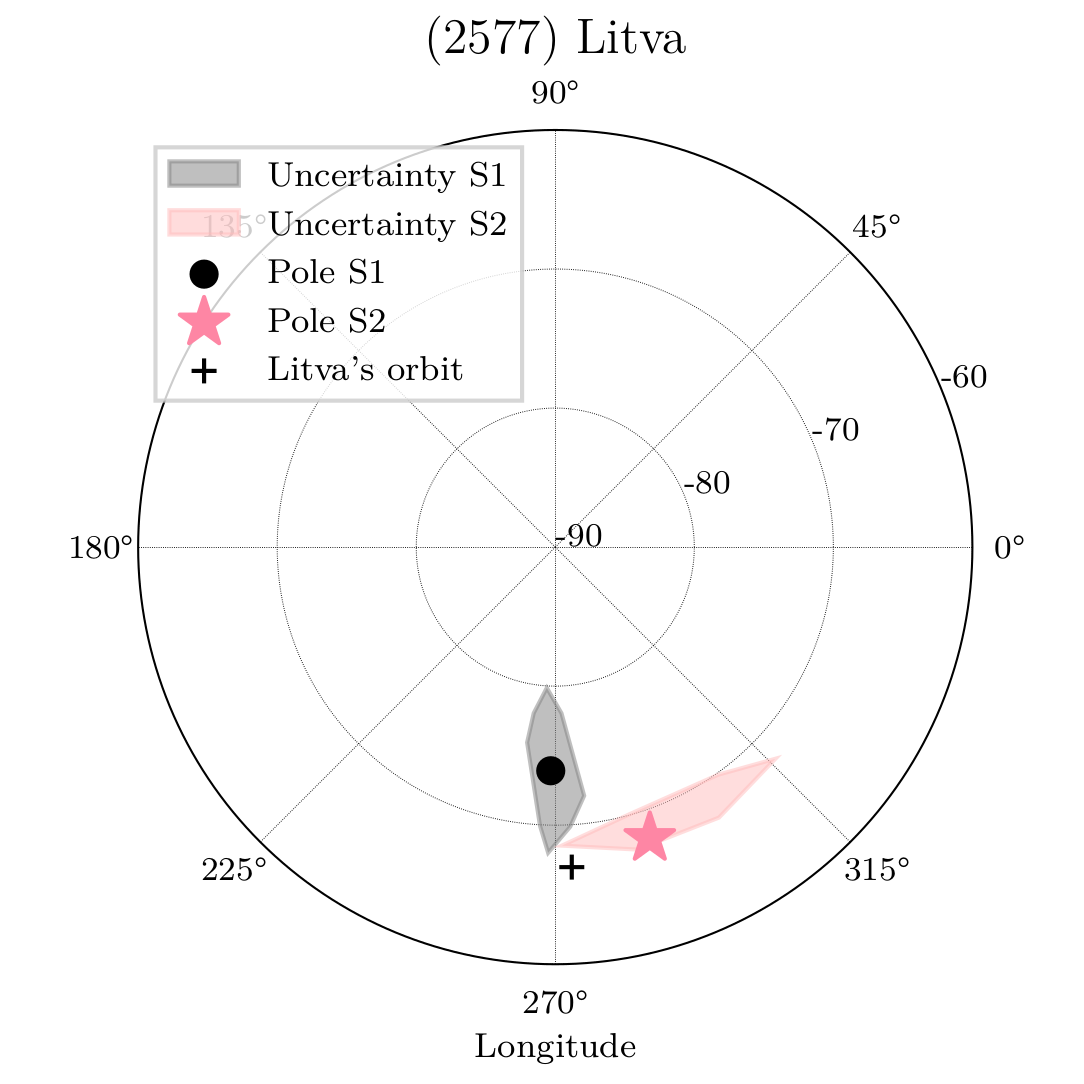}
\caption{Orbital poles of Litva's satellites with uncertainties. 3-sigma area of uncertainty of the orbital pole of Litva's inner satellite (S1) is represented in grey, and the 3-sigma area of uncertainty of the orbital pole of Litva's outer satellite (S2) in pink. The orbital pole of S1 is marked by a black circle, and the orbital pole of Litva's outer satellite S/2012 (2577) 1 (S2) is marked by a pink star. The orbital pole of Litva's heliocentric orbit is marked by a black cross. }
\label{fig:litvapole}
\end{figure}

\section{Orbit determination results}%
\label{sec:results}%

We present here the orbital solutions we found for the \glspl{vwba}.
A brief overview of all of the systems can be found in \Cref{tab:results},
and additional details for each system 
are provided 
in the following subsections. \KM{Ephemeris for these satellites, and other \genoid orbital solutions, are publicly accessible online\footnote{\url{https://ssp.imcce.fr/webservices/miriade/api/ephemsys/}\KM{Due to recent laboratory name changes, this url may change in coming years. Please contact K. Minker, B. Carry, or F. Vachier if you have trouble accessing ephemerides.}} through the LTE (previously known as IMCCE) Solar system portal (SSP).}

\subsection{(379) Huenna}
\begin{table*}
\begin{center}
  \caption[Orbital elements of the satellite of Huenna]{%
    Orbital elements of the satellite of Huenna,
    expressed in EQJ2000, obtained with \genoid.
    }
  \label{tab:huennaorbit} 
   \begin{tabular}{l ll}
    \hline\hline
    & \multicolumn{2}{c}{(379) Huenna and S/2003 (379) 1}\\ 
    \hline
  \noalign{\smallskip}
  \multicolumn{2}{c}{Observing data set} \\
  \noalign{\smallskip}
    Number of observations  & \multicolumn{2}{c}{42} \\ 
    Time span (days)        & \multicolumn{2}{c}{6671} \\ 
    RMS (mas)               & \multicolumn{2}{c}{5.45} \\ 
    \hline
  \noalign{\smallskip}
  \multicolumn{2}{c}{Orbital elements EQJ2000} \\
  \noalign{\smallskip}
    $P$ (day)         & 80.21585 & $\pm$ 0.01450 \\ 
    $a$ (km)          & 3488.8 & $\pm$ 48.4 \\ 
    $e$               & 0.283 & $\pm$ 0.010 \\ 
    $i$ (\degr)       & 151.1 & $\pm$ 0.7 \\ 
    $\Omega$ (\degr)  & 204.4 & $\pm$ 1.3 \\ 
    $\omega$ (\degr)  & 278.8 & $\pm$ 1.5 \\ 
    $t_{p}$ (JD)      & 2452930.95866 & $\pm$ 0.30025 \\ 
    \hline
  \noalign{\smallskip}
  \multicolumn{2}{c}{Derived parameters} \\
  \noalign{\smallskip}
    $M$ ($\times 10^{17}$ kg)      & 5.229 & $\pm$ 0.218 \\ 
    $\lambda_p,\,\beta_p$ (\degr)  & 161, -78 & $\pm$ 4, 1 \\ 
    $\alpha_p,\,\delta_p$ (\degr)  & 114, -61 & $\pm$ 1, 1 \\ 
    \hline
  \end{tabular}
  
  \footnotesize{ Orbital period $P$, semi-major axis $a$,
    eccentricity $e$, inclination $i$,
    longitude of the ascending node $\Omega$,
    argument of pericenter $\omega$, time of pericenter $t_p$.
    The number of observations and RMS between predicted and
    observed positions are also provided.
    Finally, we report the mass of Huenna $M_{\textrm{Huenna}}$,
    the ecliptic J2000 coordinates of the orbital pole
    ($\lambda_p,\,\beta_p$), 
    the equatorial J2000 coordinates of the orbital pole
    ($\alpha_p,\,\delta_p$).  
    Uncertainties are given at the 1-$\sigma$ level.}
\end{center}
\end{table*}

The orbital solution for asteroid (379) Huenna presented here is
a minor update to the \cite{2022Icar..38215013V} solution, \KM{yielding similar results with a significantly longer orbital baseline therefore validating the long term stability of the \cite{2022Icar..38215013V} solution}.
We find an orbital period of 80.2\,d, a semi-major axis of 3489\,km,
and and eccentricity of 0.28.
New observations on 2021-11-18 (Keck/NIRC2, PID:C232, PI:Brown) extend the dataset by an additional seven years compared to \cite{2022Icar..38215013V}, to a total of 18 years of observations. The changes in determined orbital elements are negligible given the uncertainties. Our new orbital solution, based on 42 observations spanning 6671 days, with a root mean square (RMS)
of the residuals between the observed positions and those predicted 
of 5.45 mas only ($\approx 0.5$ px), can be seen in \Cref{tab:huennaorbit}.
The significantly non-zero eccentricity observed in Huenna's satellite is similar to some
small \gls{vwba} systems,
such as (2577) Litva 
(see below).
We observe a non-negligible influence of solar perturbations.
This orbital solution is stable enough to be used reliably for
predicting stellar occultations of the satellite in the 
forthcoming years.

\subsection{(2577) Litva}
\label{sec:litvaorb}

\begin{table*}
  \caption{Table of orbital properties calculated for VWBA systems. 
  }
  \centering
\begin{tabular}{llrrrrrrrrr}
\hline\hline
  Number & Name & $a (km)$ & $\sigma_a$ & $e$ & $\sigma_e$ & $M_p (10^{14}kg)$ & $\sigma_{M_p}$ & $\rho$ & $\sigma_{\rho}$ & $D_s (km)$\\
\hline
  1509 & Esclangona &  &  &  &  & & & & & 1.4\\
  2577 & Litva & 348 & 44 & 0.30 & 0.09 & 0.72 & 0.24 & 1.9 & 1.3 & 1\\
  3749 & Balam &  & & > >0 &  & &  & & & 1\\
  4674 & Pauling & 310 & 30 & 0.035 & 0.08 & 1.33 & 0.36 & 2.4 & 0.7 & 1\\
  22899 & Alconrad & 204 & 18 & 0.13 & 0.1 & 2.13 & 0.56 & 2.9 & 0.9 & 1.8\\
  17246 & Christophedumas & 259 & 8 & 0.183 & 0.024 & 1.78 & 0.14 & 3.3 & 0.9 & 1.6 \\
  379 & Huenna & 3489 & 48 & 0.28 & 0.01 & 5230 & 220 & 1.5 & 0.1 & 3 \\
  3548 & Eurybates & 2350 & 40 & 0.13 & 0.03 & 1500 & 90 & 1.1 & 0.3 & 0.8\\
\hline 
\end{tabular}

  For (3548) Eurybates 
  the satellite diameter
  is extracted from \cite{2020MPECQuetaEurybatesDiscoNoll}. In the case of Balam and Litva, $D_s$ refers to the outermost satellite. 
\label{tab:results}
\end{table*}


\begin{table*}[h!]
\begin{center}
  \caption[Orbital elements of the satellite of Litva]{%
    Orbital solutions for the two satellites of Litva,
    expressed in EQJ2000, obtained with \genoid, and by the methods described in sections \ref{sec:lc} and \ref{sec:litvaorb}. 
    }
  \label{tab:litvaorbit} 
   \begin{tabular}{l ll}
    \hline\hline
    & \multicolumn{2}{c}{(2577) Litva and S/2012 (2577) 1}\\ 
    \hline
  \noalign{\smallskip}
  \multicolumn{2}{c}{Observing data set} \\
  \noalign{\smallskip}
    Number of observations  & \multicolumn{2}{c}{22} \\ 
    Time span (days)        & \multicolumn{2}{c}{3522} \\ 
    RMS (mas)               & \multicolumn{2}{c}{3.09} \\ 
    \hline
  \noalign{\smallskip}
  \multicolumn{2}{c}{Orbital elements EQJ2000} \\
  \noalign{\smallskip}
    $P$ (day)         & 214.79901 & $\pm$ 37.10072 \\ 
    $a$ (km)          & 347.8 & $\pm$ 43.6 \\ 
    $e$               & 0.302 & $\pm$ 0.094 \\ 
    $i$ (\degr)       & 172.9 & $\pm$ 7.4 \\ 
    $\Omega$ (\degr)  & 87.3 & $\pm$ 26.1 \\ 
    $\omega$ (\degr)  & 162.5 & $\pm$ 36.5 \\ 
    $t_{p}$ (JD)      & 2456105.35556 & $\pm$ 12.06249 \\ 
    \hline
  \noalign{\smallskip}
  \multicolumn{2}{c}{Derived parameters} \\
  \noalign{\smallskip}
    $M$ ($\times 10^{13}$ kg)      & 7.227 & $\pm$ 2.477 \\ 
    $\lambda_p,\,\beta_p$ (\degr)  & 288, -66 & $\pm$ 20, 3 \\ 
    $\alpha_p,\,\delta_p$ (\degr)  & 357, -83 & $\pm$ 26, 7 \\ 
    $\Lambda$ (\degr)              & 11.5 & $\pm$ 1.3 \\ 
    \hline
        & \multicolumn{2}{c}{Inner satellite}\\ 
    \hline
    \noalign{\smallskip}
    $P_{orb}$ (hours) & 35.87101 & $\pm$ 0.00022 \\
    $P_{orb}$ (days) & 1.4946254 & $\pm$ 0.000092 \\
    $a/D_{\textrm{eff}}$ & 2.44 & $\pm$ 0.57 \\
    $a$ (km) & 10 & $\pm$ 3 \\
    $e$ & $<$ 0.07 & \\
    $D_s/D_p$ & 0.34 & $\pm$0.02 \\
    $c/a$ & 0.67 & +0.33 -0.23 \\
    $\lambda_p,\,\beta_p$ (\degr)  & 269.0, -74 & $\pm$ 7.3, 6 \\
    \hline \\
  \end{tabular}

  \footnotesize{Please see Table \ref{tab:huennaorbit} for a description of the listed parameters.
    Uncertainties are given at 1-$\sigma$ for the outer satellite, and 3-$\sigma$ for the inner satellite. 
    $D_{\textrm{eff}}$ represents the equivalent diameter of the Litva system, and the
    orbital inclination ($\Lambda$) with respect to the orbital plane of the inner satellite.}
\end{center}
\end{table*}

For this triple asteroid system, we present two orbital solutions:
one
for the inner satellite based on photometric
lightcurve observations of the system (\cref{sec:lc})
and a second
for the outer satellite based on direct measurements
of satellite positions (\cref{ssec:images}).
Both of these solutions are described in \Cref{tab:litvaorbit}. 

The orbital solution for the inner satellite,
based on a nine-year period of observations (2009-2018),
is in general agreement with the previously published solution 
from \cite{pravec2012}.
We determine an orbital period of $35.87101 \pm 0.00022$\,h, with uncertainties at the $3 \sigma$ level.
The solution that we present includes the known outer satellite 
of $D_s/d_c=0.23$, where $d_c$ is the quadratic sum of the diameters of the primary and the inner satellite
(\Cref{app:obs:sat}).

The orbital solution derived here for Litva's outer satellite confirms
the very long orbital period reported by \cite{2013merlinelitvadiscoiauc}, and we find good agreement between the periods ($\numb{215\pm37}$ against \numb{214} days) and the 
semi-major axes (\numb{347\,$\pm$\,44}\,km against \numb{378}) of the two solutions.
This solutions encompasses
 the observations used for the preliminary study of the orbit
at the time of discovery, with 
an additional epoch in \numb{2004} providing a much longer temporal baseline
(\Cref{tab:litvagenoid}). 
The orbit is notably eccentric ($e=0.31$), which is uncommon among most
asteroid satellites, but similar to that of the satellite of (379) Huenna
\citep[see above and][]{2022Icar..38215013V}.
The positions of the satellite predicted by this orbital solution match the  observed positions very well, with RMS residuals of \numb{3.09} mas.

Our initial orbital solution for Litva, which used only the observations from 2012-2013,
was degenerate, with two possible orbital-plane-orientations for the system.
Ultimately, we were able to reject one solution by considering the
orbital orientation of the inner satellite, under the assumption that
the orbits of both satellites are not far from co-planar
\citep[as is the case for most triple systems, e.g.,][]{2011Icar..216..241B, 
2021A&A...650A.129C, 2021A&A...653A..56B}. However, a slight (10\degr)
offset can be observed between the orbital axis of the two satellites
(\cref{fig:litvapole}). 
We were then able to confirm this by comparison with the 2004 observation,
which provided a faint detection of the satellite close to the primary
and an exceptionally good match to the preliminary ephemeris, 
despite the eight year observational gap between the 2004 and 2012 observations.
This point was included in the final orbital solution.

Photometric lightcurve
observations of Litva indicate an axial ratio
$c/a$ of 0.67$_{-0.23}^{+0.33}$ (3-sigma uncertainties).
This suggests that Litva may be similarly oblate to other small,
well studied binary systems, including (65803) Didymos
\citep[$c/a=0.72$,][]{2024chabotdart} and
(66391) Moshup
\citep[$c/a=0.86$,][]{2006Ostro1999Kw4}.
In the absence of an accurate shape model of Litva's primary,
it is unclear whether Litva shares the elongated-ellipsoid 
shape common amongst larger binary systems 
\citep[for example, (243) Ida, see the trend in][]{2021A&A...650A.129C},
or the top-shape 
common amongst NEA 
binaries (for example, (65803) Didymos). 
Development of an accurate shape model for Litva or any 
other \gls{vwba} systems could provide valuable clues to their formation
\citep{2022NatCo..13.4589Z}. 
However, the relatively low lightcurve amplitude of litva (0.17 mag) across multiple apparitions suggests that the system's primary is unlikely to be heavily elongated.

\begin{table*}[h!]
\begin{center}
  \caption[Orbital elements of the satellite of Pauling]{%
    Orbital elements of the satellite of Pauling,
    expressed in EQJ2000, obtained with \genoid; please see Table \ref{tab:huennaorbit} for a description of the listed parameters.}
  \label{tab:paulingorbit} 
   \begin{tabular}{l ll}
    \hline\hline
    & \multicolumn{2}{c}{(4674) Pauling and S/2004 (4674) 1}\\ 
    \hline
  \noalign{\smallskip}
  \multicolumn{2}{c}{Observing data set} \\
  \noalign{\smallskip}
    Number of observations  & \multicolumn{2}{c}{14} \\ 
    Time span (days)        & \multicolumn{2}{c}{3055} \\ 
    RMS (mas)               & \multicolumn{2}{c}{5.13} \\ 
    \hline
  \noalign{\smallskip}
  \multicolumn{2}{c}{Orbital elements EQJ2000} \\
  \noalign{\smallskip}
    $P$ (day)         & 133.29069 & $\pm$ 0.25397 \\ 
    $a$ (km)          & 310.0 & $\pm$ 29.9 \\ 
    $e$               & 0.035 & $_{-0.035}^{+0.080}$ \\ 
    $i$ (\degr)       & 161.2 & $\pm$ 3.8 \\ 
    $\Omega$ (\degr)  & 132.9 & $\pm$ 8.9 \\ 
    $\omega$ (\degr)  & 174.4 & $\pm$ 50.1 \\ 
    $t_{p}$ (JD)      & 2453118.40620 & $\pm$ 17.68012 \\ 
    \hline
  \noalign{\smallskip}
  \multicolumn{2}{c}{Derived parameters} \\
  \noalign{\smallskip}
    $M$ ($\times 10^{14}$ kg)      & 1.329 & $\pm$ 0.367 \\ 
    $\lambda_p,\,\beta_p$ (\degr)  & 325, -73 & $\pm$ 13, 3 \\ 
    $\alpha_p,\,\delta_p$ (\degr)  & 43, -71 & $\pm$ 9, 4 \\ 
    \hline
  \end{tabular}
\end{center}
\end{table*}

\subsection{(3548) Eurybates}

\begin{table*}
\begin{center}
  \caption[Orbital elements of the satellite of Eurybates]{%
    Orbital elements of the satellite of Eurybates,
    expressed in EQJ2000, obtained with \genoid please see Table \ref{tab:huennaorbit} for a description of the listed parameters.}
  \label{tab:eurybatesorbit} 
   \begin{tabular}{l ll}
    \hline\hline
    & \multicolumn{2}{c}{(3548) Eurybates and Queta}\\ 
    \hline
  \noalign{\smallskip}
  \multicolumn{2}{c}{Observing data set} \\
  \noalign{\smallskip}
    Number of observations  & \multicolumn{2}{c}{9} \\ 
    Time span (days)        & \multicolumn{2}{c}{884} \\ 
    RMS (mas)               & \multicolumn{2}{c}{7.89} \\ 
    \hline
  \noalign{\smallskip}
  \multicolumn{2}{c}{Orbital elements EQJ2000} \\
  \noalign{\smallskip}
    $P$ (day)         & 82.5 & $\pm$ 0.2 \\ 
    $a$ (km)          & 2345.5 & $\pm$ 43.6 \\ 
    $e$               & 0.14 & $\pm$ 0.03 \\ 
    $i$ (\degr)       & 132.6 & $\pm$ 0.6 \\ 
    $\Omega$ (\degr)  & 193.5 & $\pm$ 2.5 \\ 
    $\omega$ (\degr)  & 15.7 & $\pm$ 10.4 \\ 
    $t_{p}$ (JD)      & 2458375.07879 & $\pm$ 1.88861 \\ 
    \hline
  \noalign{\smallskip}
  \multicolumn{2}{c}{Derived parameters} \\
  \noalign{\smallskip}
    $M_{\textrm{Eurybates}}$ ($\times 10^{17}$ kg)      & 1.502 & $\pm$ 0.089 \\ 
    $\lambda_p,\,\beta_p$ (\degr)  & 113, -64 & $\pm$ 4, 1 \\ 
    $\alpha_p,\,\delta_p$ (\degr)  & 104, -43 & $\pm$ 3, 1 \\ 
    \hline
  \end{tabular}
\end{center}
\end{table*}

Eurybates, future rendezvous target of NASA's Lucy mission, is the only known VWBA system among the Jupiter Trojans. Previous solutions have been published by \cite{2021-Brown-Eurybates} and \cite{Nolleurybates2023}, with a minor disagreement ($3\sigma$) in eccentricity between the two. Here, we present an alternative solution calculated from the astrometry presented in \cite{2021-Brown-Eurybates}. The \cite{Nolleurybates2023} solution also contains additional HST observations from December \KM{2021 (mistakenly reported as December 2022 in \cite{Nolleurybates2023})},  
For consistency, we choose to only include the astrometric positions reported by \cite{2021-Brown-Eurybates}. We reproduce these positions in \Cref{tab:eurybatesgenoid}, with minor corrections to astrometric sign errors and observation dates.

The results of our solution are presented in \Cref{tab:eurybatesorbit}. We use a model that includes Keplerian motion and Solar tides, which produced results consistent with those of \cite{2021-Brown-Eurybates} within $1 \sigma$ uncertainties. The inclusion of Solar tides, \textcolor{black}{which were included in the Noll solution but not the Brown solution,} could explain the discrepancy between the \cite{2021-Brown-Eurybates} and \cite{Nolleurybates2023} models. We note that the statistical uncertainties present by \genoid are often somewhat overestimated for certain parameters, \KM{and that our Solar-tide model matches slightly better with the tideless Brown model than the Solar-tide Noll model}. 

We also attempted a solution with fixed astrometric uncertainties for all points (as was done with HST solutions for Alconrad and Christophedumas), rather than the astrometric uncertainties presented by \cite{2021-Brown-Eurybates}. This presented a slightly better fit (RMS=\numb{6.38}\,mas vs. RMS=\numb{7.89}\,mas).

\subsection{(3749) Balam}

Balam is one of the most interesting known multiple asteroid systems, exhibiting a complex dynamical geometry involving one close satellite, one distant satellite, and one unbound (escaped) satellite, in the form of a pair \citep{2019pravecpairs}. Until this point, (3749) Balam has been the only small \gls{vwba} system for which
an orbital solution has been published in the literature
\citep{marchis2008eccentric}, 
albeit the solution was heavily degenerate
and numerous sets of parameters could fit the data equally well
\citep[see][]{2012AA...543A..68V}.
However, as pointed out in \Cref{sec:Observations}, the field orientation 
of the earliest Gemini/Hokupa'a data is deemed unreliable.
This significantly
reduces the amount of available positions to 
9 observations over only 17 days.
As the orbital period could be over a factor of four longer than the time span of the observations, 
this extremely limited dataset prohibits the development of a meaningful orbital solution.

We note that eccentric solutions were slightly more favorable than strictly circular solutions (9.8 mas residuals vs 11.2 mas residuals), with reasonable solutions found for $0.2<e<0.8$. The lack of detection of the satellite 
in some images
suggests that the satellite might be very close to the primary at this time, supporting this hypothesis. Notably, the prior orbital solutions presented significant degeneracies in key parameters \citep[see][]{vachier2012genoide} and large uncertainties, supporting our decision to exclude these observations.

\subsection{(4674) Pauling}

For (4674) Pauling, no previously-computed orbital solution 
can be found in the literature. Our solution
fits the \numb{14} observations with a RMS residual of \numb{5.13}\,mas.
The semi-major axis is found to be \numb{310}\,$\pm$\,\numb{30}\,km,
and the low eccentricity
($e = 0.04^{+0.08}_{-0.04}$) is compatible with a purely circular orbit.
Assuming a primary diameter of 4.7\,$\pm$\,0.1 km (\cite{{2022MyhrvoldWISE}}, although diameters from mid-infrared
radiometry may be affected by systematic errors of about
10\%, \cite{2018AJ....156...62M})
and similar densities for the primary and satellite,
we determine a density of 2.4\,$\pm$\,0.7 g.cm$^{-3}$ for the system.
The value is in line with density values reported for other small
S-type asteroids \citep[e.g.,][]{2012PSS...73...98C, 2022PSJ.....3..163S}.
A detailed description of the orbital solution can be found in \Cref{tab:paulingorbit}.

\subsection{(17246) Christophedumas}
\label{sec:cdorb}

Here we present our orbital analysis for (17246) Christophedumas.
Unfortunately, there is a high level of ambiguity in this orbital solution, \KM{as such we were not able to achieve an} orbital fit that is physically reasonable and also has an acceptable RMS. 
We were able to develop an extremely low-residual (0.23 mas RMS)
orbital solution for the system. The physical parameters
derived from this solution are, however, in contrast
to our expectations.
The orbital solution (in combination of the system diameter of $\numb{4.7\pm0.4}$\,km determined in section \ref{sec:CDdiam}) indicates an extremely high-density primary,
($\numb{6.6}\pm1.8$ g.cm$^{-3}$), which poorly aligns with 
other members of the
Koronis family \citep{1997PetitIda}, other \gls{vwba} systems
(see \cref{tab:results}),
and the general asteroid population \citep{2012PSS...73...98C, 2021A&A...654A..56V}.
Furthermore, this orbital solution indicates that the satellite should be
at a wider separation from the primary than any other epoch 
of observation during the 2005-07-04 non-detection. 
Since there are no obvious issues with image quality,
it is suspicious that the satellite would not be resolvable at this time,
given that it was successfully resolved at much smaller angular separations. \KM{Due to the algorithm used by \genoid to determine determine best-fitting parameters, inserting a ``false'' point with a very large uncertainty to represent the non-detection would heavily bias the results, and as such it is not possible to constrain the solutions in this way.} Therefore, we discard this solution.

We were able to determine an alternate solution by restricting 
the mass of the system to only provide solutions with a physically
reasonable ($\rho\lesssim\numb{4}$ g.cm$^{-3}$) density. 
The orbital fit is substantially worse (6.4 mas RMS), but the non-detection is more reasonably placed. This is significantly worse than the precision expected from HST observations, the precision achieved for (22899) Alconrad, based on a very similar dataset, and the precision of many solutions based on ground based observations. However, in light of the smearing issue affecting one in four images, the fit could be acceptable in this instance. This solution provides an eccentricity of ($e\approx0.18$), and a density of $\numb{3.3\pm0.9}$ g.cm$^{-3}$, \KM{however, the extremely small and poor-quality dataset for Christophedumas means that these results are merely the best approximation possible with the current, very limited dataset, and would benefit significantly from future follow-up observations.}
All reasonable solutions we have determined require non-zero eccentricity of the satellite.

\subsection{(22899) Alconrad}

For (22899) Alconrad, we were able to determine an unambiguous orbital
solution for the system. From 5 observations, we find an orbital period of 56\,days,
a semi-major axis of 204\,$\pm$\,18\,km and 1.46 mas RMS residuals.
Taking the diameter of \numb{5.2\,$\pm$\,0.3}\,km
(\Cref{sec:physical}),
we determine a density of $\numb{2.9\pm0.9}$ g.cm$^{-3}$. This is consistent with the density of Ida and other members of the Koronis family.
A detailed description of the orbital solution is reported in  \Cref{tab:alconradorbit}.
\begin{table*}[h!]
\begin{center}
  \caption[Orbital elements of the satellite of Alconrad]{%
    Orbital elements of the satellite of Alconrad,
    expressed in EQJ2000, obtained with \genoid; please see Table \ref{tab:huennaorbit} for a description of the listed parameters.
    }
  \label{tab:alconradorbit} 
   \begin{tabular}{l ll}
    \hline\hline
    & \multicolumn{2}{c}{(22899) Alconrad and Juliekaibarreto}\\ 
    \hline
  \noalign{\smallskip}
  \multicolumn{2}{c}{Observing data set} \\
  \noalign{\smallskip}
    Number of observations  & \multicolumn{2}{c}{5} \\ 
    Time span (days)        & \multicolumn{2}{c}{520} \\ 
    RMS (mas)               & \multicolumn{2}{c}{1.46} \\ 
    \hline
  \noalign{\smallskip}
  \multicolumn{2}{c}{Orbital elements EQJ2000} \\
  \noalign{\smallskip}
    $P$ (day)         & 56.29374 & $\pm$ 0.14665 \\ 
    $a$ (km)          & 204.2 & $\pm$ 18.0 \\ 
    $e$               & 0.132 & $\pm$ 0.100 \\ 
    $i$ (\degr)       & 149.2 & $\pm$ 2.0 \\ 
    $\Omega$ (\degr)  & 95.4 & $\pm$ 3.0 \\ 
    $\omega$ (\degr)  & 251.5 & $\pm$ 15.0 \\ 
    $t_{p}$ (JD)      & 2452847.00556 & $\pm$ 1.92780 \\ 
    \hline
  \noalign{\smallskip}
  \multicolumn{2}{c}{Derived parameters} \\
  \noalign{\smallskip}
    $M$ ($\times 10^{14}$ kg)      & 2.128 & $\pm$ 0.561 \\ 
    $\lambda_p,\,\beta_p$ (\degr)  & 331, -54 & $\pm$ 3, 2 \\ 
    $\alpha_p,\,\delta_p$ (\degr)  & 5, -59 & $\pm$ 3, 2 \\ 
    \hline
  \end{tabular}
\end{center}
\end{table*}

\section{Discussion}%
\label{sec:discussion}%

\subsection{Proposed formation mechanisms of \glspl{vwba} }

In the following, we will discuss various mechanisms for \gls{vwba} formation that
have been proposed by previous authors, including escaping ejecta binaries (EEBs),
Binary-YORP (BYORP) spreading, and alternative YORP mechanisms. Following these descriptions,
we will discuss the applicability of these models to each of the \gls{vwba} systems studied in previous section.

\subsubsection{Escaping Ejecta Binary model}

The Escaping Ejecta Binary (EEB) model induces binary formation through mutual
capture of two (or more) ejecta particles following an impact to a shared 
parent body \citep{2004Icar..167..382D}. 
The chaotic nature of this formation mechanism results in a range of
orbital parameters, and predicts a wide range of eccentricity

\citep{2004Icar..167..382D}.

Depending on the specifics of the system, it is possible that eccentricity
could be damped over time due to tidal dissipation \citep{GoldreichSari2009TidalEvolutionBinaries}.
Simulations by \citet{2004Icar..167..382D} suggest that a large fraction 
of EEBs could have equal size ratios ($D_s/D_p>0.5$), however this is most likely an
effect of the minimum particle size used in the simulations.
The nature of this formation mechanism should favor systems that belong to
(large) families, as the same impacts that form these families would
produce the ejecta necessary for the formation of these binary systems. 
It is also possible that the very wide satellites formed as a result of an impact to the primary body of the system (analogous to the SMATs of the \citet{2004Icar..167..382D} model). This is particularly likely for Eurybates, which is the largest member of the Eurybates family. However, we consider the EEB and SMAT formation scenarios proposed by \cite{durda2004EEB} to be roughly equivalent, as in the limit of catastrophic disruption where there are several equally large family members instead of a single parent-body remnant there is no meaningful difference between the two. As such, we henceforth refer to these models jointly as the "EEB model", although we note that this is indeed an oversimplification.

\subsubsection{BYORP expansion}

The very wide binary asteroid population has also been previous linked to YORP spin-up,
a formation mechanism which is expected to have produced the majority of known
small ($D<10$\,km) binary asteroid systems,
as several members of this population exhibit characteristic YORP formation traits
\citep[$D_s/D_p\approx0.3$, spin periods $\approx$2.5\,h,][]{2011Icar..212..167P, 2007Icar..190..250P}.
Binary-YORP (BYORP) effects can influence the orbits of asteroid satellites,
causing them to expand or contract, and gain or lose eccentricity \citep{2014jacobsonVWBAs}.

It is possible that the \glspl{vwba} initially form their satellites at close distances
($a\approx 3 R_p$), and then BYORP effects expand the orbit to their currently observed very wide positions \citep{2014jacobsonVWBAs}.
Stable BYORP expansion of an asteroid satellite formed through typical YORP spin-up could present a very circular orbit that is co-planar to the spin orientation of the primary,
so that the expansion remains stable over time 
\citep{CukandBurns2005BYORP}.
BYORP expansion also requires a tidally locked satellite,
although systems that were previously expanded by BYORP may have satellites that are no
longer tidally locked \citep{2014jacobsonVWBAs}. There is also a possibility that in some systems BYORP expansion occurs faster than tidal circularization, which could potentially result in a non-zero eccentricity for an expanded system. Similarly, solar tides, which become stronger at greater separations, can act to grow the eccentricity over time, particularly if the heliocentric orbit itself becomes more eccentric. As such, while a perfectly circular orbit for a distant satellite would be a marker of BYORP expansion, the lack thereof does not exclude the possibility of previous BYORP expansion.

Previous studies \citep{2014jacobsonVWBAs, polishook2011vwba} have argued against the EEB formation scenario in favor of a BYORP scenario for small VWBAs as these systems typically possess rapidly rotating primaries (see \Cref{tab:general}), which are consistent with YORP spin-up. However, rapidly rotating primaries are characteristic of small mafic-silicate rich asteroids in general (see \Cref{fig:spungarias}), and as such this is insufficient to exclude the EEB scenario. A potential marker of the YORP satellite formation could be an alignment between the satellite's orbital pole and the primary's spin axis, although this may not occur in the case of multiple-satellite interaction. Nevertheless, the observational dataset is currently insufficient to determine the primary spin axis to the necessary precision for most systems, so it is not possible to test. It is also worthy to note that the two scenarios are not mutually exclusive, as a satellite formed through the EEB scenario could later undergo the influences of BYORP, given the appropriate orbital geometry.

\subsubsection{Simultaneous YORP formation and scattering}

It is also possible that the \gls{vwba} population forms
through the typical YORP formation pathway with closely orbiting satellites
that are then expanded to their current very wide orbits by a mechanism other than BYORP.
One dynamical explanation for this could be that in systems with multiple satellites,
mutual interactions may occur that scatter
one or more of the satellites onto a very wide orbit.

This mechanism is also capable of ejecting one (or more) satellites from the system,
forming an asteroid pair.
The scattering event could position the satellite at a random inclination or
eccentricity, resulting in a wide variety of final orbital geometries.
This argument is particularly compelling in systems with multiple known
satellites, although it could be consistent with single-satellite systems.
For single-satellite systems, it is possible that one satellite is ejected
from the system, or that there is an unknown additional satellite orbiting
on a moderately-wide orbit that our current detection methods are not robust to.
This mechanism is briefly described in \cite{Agrusa2024DART},  as well as \cite{2019pravecpairs}, which notes an overlap between asteroid pair primaries and known (Balam) or suspected very wide satellites. A potential marker of this mechanism would be that the pericenter of the outer satellite be just outside the orbit of the inner satellite, although this is not observed for any of the systems in this study.

\subsection{Comparison with observed properties}

In general, we find the orbital dynamics of the \glspl{vwba} reported in 
\Cref{sec:results} to be largely incompatible with active BYORP expansion. 
Of the four small ($D<10$\,km) systems for which we were able to develop orbital models,
all but Pauling were incompatible with a zero-eccentricity model.
Since active BYORP expansion requires the satellite to be on a circular equatorial orbit,
these non-zero eccentricities indicate that another mechanism must be fully or partially
responsible for the current orbital orientation. 

Furthermore, the wide satellites of both Litva and Esclangona are known to have spin periods
independent of both the orbital period and primary spin period \citep{2011WarnerLitvaLC3rdperiod, 2009esclangonawarner},
prohibiting them from maintaining the tidally locked singly-synchronous geometry
required to maintain active BYORP expansion \citep{CukandBurns2005BYORP}.
For both systems the satellite spin period is very close to twice the
primary spin period, varying by a factor of 2.02 and 2.04 respectively.
This notable correlation suggests that there may be some kind of ongoing process
or formation artifact stabilizing the spins of these outer satellites,
or potentially that these spin periods have been misattributed
to the outer satellite, and actually belong to some other component of the system.
The latter may imply that the fraction of \glspl{vwba} is quite large,
as this third spin period component was used to predict the discovery of Litva's
outer satellite \citep{pravec2012}. Furthermore, the successful discovery of a third component based on a misattributed
spin period implies that there is a high likelihood of any given binary
system containing an additional component \citep{2011WarnerLitvaLC3rdperiod}.

Based on these two pieces of evidence, we determine that the following systems are
unlikely to have formed through YORP spin-up followed by BYORP expansion:
(1509) Esclangona, (2577) Litva, (22899) Alconrad, and (17246) Christophedumas.
The satellite of (4674) Pauling could plausibly have formed through this scenario,
and the orbital solution for (3749) Balam is too poor in quality to make an accurate
determination.
For these systems, two possibilities remain: the EEB model or an exotic YORP formation mechanism (for example, mutual interactions between satellites, one of which may currently be unbound). 
For the two large VWBA systems, (379) Huenna and (3548) Eurybates,
the EEB model seems most likely.
Both systems are associated with large families (Themis and Eurybates),
indicating the presence of a major impact in their dynamical history.
Both systems exhibit significantly eccentric orbits,
aligning with the somewhat random orbital dynamics expected for EEB satellites.
Furthermore, YORP spin-up is ineffective for large asteroids,
and both systems rotate much slower than necessary for YORP induced satellite formation
 \citep[7 or 14.1 h for Huenna and 8.7 h for Eurybates][]{2023MNRAS.521.3925X, 2010MPBuWarnerHuenna, 2021MPBu...48...13S}.
Compared to large Main-Belt and Trojan binary systems that are assumed to have
formed satellites through large impacts
such as (22) Kalliope, (702) Alauda, (283) Emma, (624) Hektor,
these slower rotation periods are longer than average, 
but within the observed range \citep{2021A&A...650A.129C},
further indicating the likelihood of impact formation.

The EEB model is very flexible, and encompasses systems resulting in a wide variety
of orbital orientations and component size ratios. 
Unfortunately, this binary asteroid formation scenario has not been studied
rigorously outside of the characterization of the Koronis family by \cite{2004Icar..167..382D},
so it is difficult to assess the exact compatibility of the mechanism
with the observed characteristics of VWBA systems.
Most known \gls{vwba} systems belong to large families, which aligns with the EEB model.
It is, however, unclear to what extent the expectation of binary systems to belong to
families has introduced an observational bias to the discovery of these systems.
For example, Alconrad and Christophedumas were both discovered as part of a
targeted study of the Koronis family.
The component size ratios ($D_s/D_p$) of most small \gls{vwba} systems is similar
to that of most binary systems that have formed through YORP spin-up, but it is unclear to what extent this is an observational bias.
Direct imaging of small binary asteroid systems already poses a significant challenge, and it is possible that a small system with the extreme component size ratios observed in large ($D>50$\,km) \gls{vwba} systems is beyond our current detection limits. 

\begin{figure}[t]
\centering
\includegraphics[width=\columnwidth]{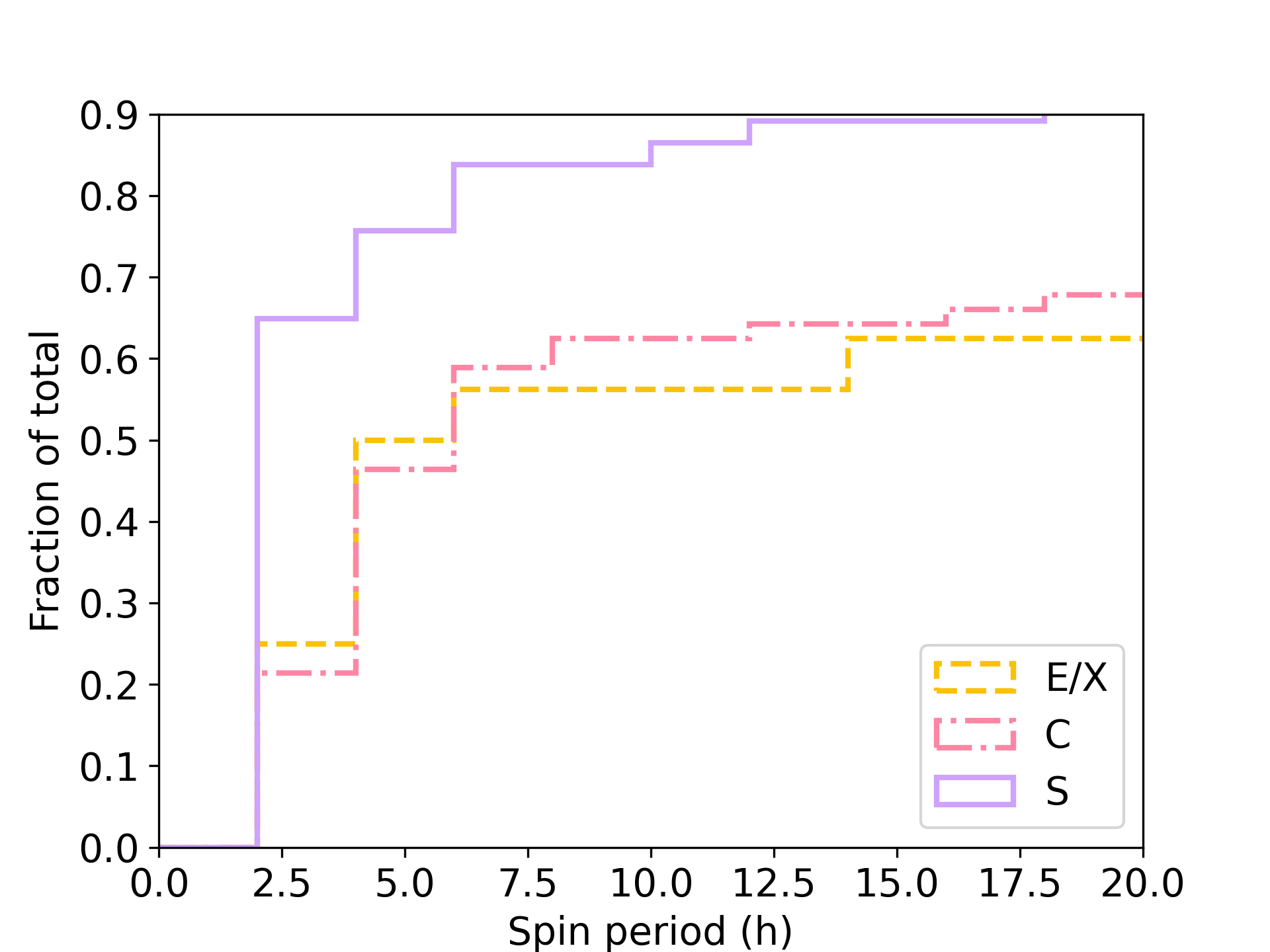}
\caption{Cumulative histogram of spin rates for Hungaria family asteroids
  across different taxonomic types, expressed as a fraction of the total population. Asteroids classified as X-types are assumed to be E-types, and have been grouped together. 
  A total sample of 39 (65, 16) S (C, X/E)-type asteroids
  with known spin periods were considered. Very slow rotating asteroids (P>100\,h) were excluded from the sample. S-type asteroids tend to rotate significantly
  faster than C or X/E type asteroids. The sample is diameter limited (2.5\,km$<D<$10\,km), with diameters estimated based on the absolute magnitude and average albedo for the taxonomic class \citep[from][]{2022MaxTax}. Analysis without the diameter limitation and sample sizes of 88 (106, 167) S (C, X/E)-type asteroids provide similar results.
  }
\label{fig:spungarias}
\end{figure}

There remain several points of evidence linking \gls{vwba} systems to some
sort of YORP formation: many systems have very fast primary spin periods,
and the $D_s/D_p$ ratio is in line with that of the population of closely-orbiting
binaries of YORP formation, although this may be influenced by observational biases.
These similarities could imply that a more exotic YORP mechanism involving
multiple satellites and/or pair formation is the source of \gls{vwba} systems. 
However, YORP can spin up an object without forming a satellite, 
\citep{2022NatCo..13.4589Z}
so a certain level of skepticism should be applied to this thinking.
This is taxonomically linked, as dark taxonomic types are more likely to deform under slower rotations.
YORP spin-up seems to be very effective on small, S-type asteroids,
causing S-type objects to reach much faster rotation rates than
C- or E-type objects in the same dynamical populations
(\Cref{fig:spungarias}).

This likely contributes to the general overabundance of silicate compositions amongst YORP-induced binary systems \citep{2023A&A...672A..48M},
but the correlation in spin rates amongst taxonomic types can also be seen in the general asteroid population.
Due to the limiting magnitude of adaptive-optics instruments,
small asteroids of dark taxonomic types are necessarily impossible to observe under normal circumstances. As such, no small \gls{vwba} systems with dark taxonomic types have been identified, although dark taxonomic types dominate the large \gls{vwba} systems.
It is impossible to know if the absence of \gls{vwba} systems amongst dark taxonomic types is a true quality of the population or merely an artifact of observational bias.

We find Litva to be notably less dense than Pauling and other small S-type asteroids \citep{2024chabotdart, 2012-PSS-66-Carry}. The extremely high albedo of the system \citep[0.56,][]{2017ali-lagoaMarsCrosserDiameters}{}{} suggests that it is unlikely that this low density results from an overestimate of the diameter of the system. It is possible that this value for Litva's albedo is overestimated (see alternative $H_V$ estimate in \cite{2012Icar..221..365P}), but correcting for this Litva is still brighter than average for an S-type asteroid ($p_V=0.41\pm0.08)$.
Spectral analysis \citep{2022MaxTax}
confirms that Litva is an S-type asteroid, rather than a mis-classified E-type. Extremely high albedos can be observed in other Hungaria-family asteroids with clear silicate features in their spectra, such as (1727) Mette.

\begin{figure}
    \centering
    \includegraphics[width=\linewidth]{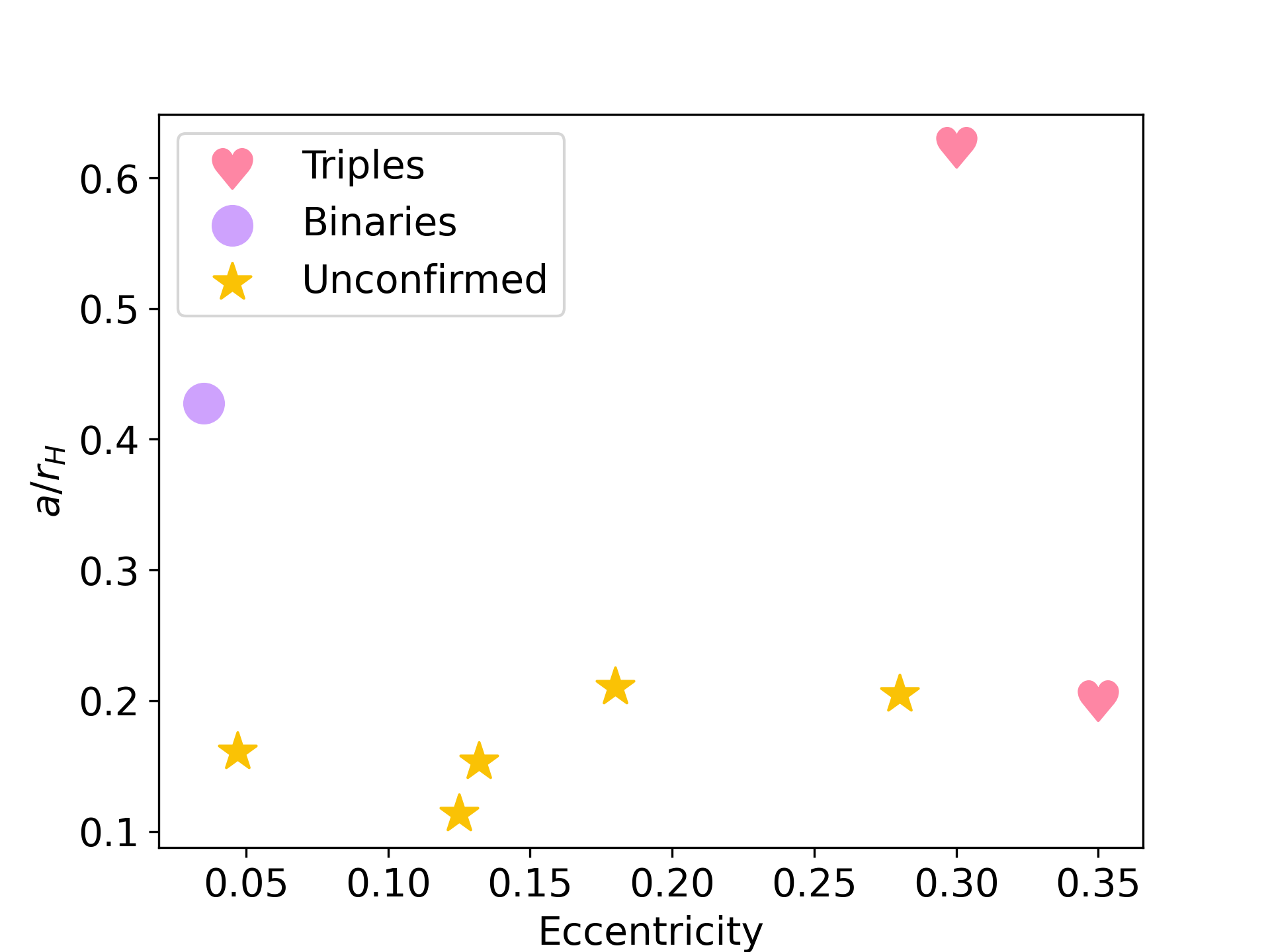}
    \caption{Normalized semi-major axis vs. eccentricities of VWBA systems. Systems with confirmed inner satellites are very eccentric, whereas those with only one satellite are nearly circular. Objects marked 'unconfirmed' are those for which current observations are insufficient to detect a satellite of similar size to the known satellite orbiting close to the system's primary. Both solutions for Christophedumas are represented, and the eccentricity and semi-major axis of Balam is approximated based off preliminary orbital solutions. Satellite eccentricity appears to be independent of semi-major axis.}
    \label{fig:eccentricities}
\end{figure}

Among small VWBA systems, we notice a correlation between the number of satellites possessed by a system and the eccentricity of the system's wide satellite. The only object in this study was an eccentricity compatible with a perfectly circular orbit, (4674) Pauling, is also the only system with enough lightcurve observations to indicate the absence of an inner satellite \citep{2006MPBu...33...34W, 2011MPBu...38...25W}. (2577) Litva has an inner satellite and a high eccentricity, as does (3749) Balam (although in the latter case this eccentricity is poorly defined). Balam is also a member of a pair, which can be considered as an escaped satellite with eccentricity >1. This indicates the possibility that the inner satellite may be influencing the eccentricity of the outer satellite, and could therefore indicate an undetected inner satellite for other eccentric VWBA systems, such as (379) Huenna. A summary of satellite eccentricities for the VWBAs can be found in Figure \ref{fig:eccentricities}. Although we note a trend between multiplicity and high eccentricities, the relationship is not necessarily causal, but is worth exploring in future work.

This spin period ratio also denotes a contrast to the wide binary systems discussed in various works by Warner et al. \citep[][to name a few]{2020MPBu...47...37W, 2019WarnerTrioOfVWBA, 2016MPBu...43..306W}, which propose a sample of objects with long (10s-100s of hours) primary spin periods and short secondary spin periods, marking a binary system with a primary which has been spun-down by YORP. This is inconsistent with the sample we present, all of which have fast primary spin periods, with the exception of Christophedumas, for which the spin period is uncertain. However, some of the proposed wide binary systems are consistent with the remote sensing properties of (152830) Dinkinesh, for which the dominant 53\,h period presented by \cite{Mottoladinkinesh} is actually associated to the bi-lobated satellite and not the primary, as initially assumed. It may be worthwhile to consider the possibility that these objects could have extremely elongated, tidally locked satellites (corresponding to the higher-amplitude, slower period) and nearly-spherical primaries (corresponding to the lower-amplitude, faster period) causing them to present this unusual profile.

\subsection{Limitations}

With the exception of (379) Huenna,
the orbital data sets in this study are sparse, 
containing few observations with large gaps between them. This allows for potential aliasing in orbit determination, and limits the precision of the presented solutions.

We encourage future observations of these systems to help resolve
these degeneracies and improve the quality of these orbits,
particularly in the case of Balam, which is a rare triple+pair system. 
Although these objects have not been observed in many years, these gaps in the observational history should be relatively easy to reconcile due to the extremely long orbital periods of the systems.

\section{Conclusions}%
\label{sec:conclusions}%

We present observations and orbital analysis
of \num{7} multiple systems of asteroids belonging to the
class of \glsdesc{vwba}s, of which we were able to develop an orbital solution for \num{5} systems.
We gather and report 
high-angular resolution images
of each system, mid-infrared photometry for
(17246) Christophedumas and (22899) Alconrad, 
and optical lightcurves for (2577) Litva.
For all systems except Christophedumas, Esclangona, and Balam, we constrain the orbit of the satellite with high accuracy. We report orbital solutions
for the first time for Pauling, Alconrad and Christophedumas, and improve or confirm the dynamics of
Litva, Huenna, and Eurybates. We also reject the previously published solution of Balam's outer satellite, but are unable to offer a new, more accurate solution with the available archival data. 
The analysis of the mid-infrared photometry
provides the diameter and albedo of Alconrad and Christophedumas.
We report the density of Alconrad, Litva, and Pauling for the first time.
Based on the properties of the systems, we
determine that only (4674) Pauling could be compatible with
the BYORP expansion model of \gls{vwba} formation.
Other systems in this sample could be compatible with a more exotic
YORP formation mechanism or an EEB model.
An apparent correlation between systems with at least two satellites
and highly eccentric orbits implies that an inner satellite can play
a role in shepherding the eccentricity of the very wide outer satellite,
a mechanism that could also apply to large-diameter very wide systems
(379) Huenna and (3548) Eurybates.

\section*{Acknowledgements}%
\label{sec:acknowledgements}%
We wish to thank Harrison Agrusa for his very helpful commentary. The work by P.S. and P.P. was supported by the Grant Agency of the Czech Republic, Grant 23-04946S. M.H. thanks the Slovak Grant Agency for Science VEGA (grant No. 2/0059/22) and M. Pikler for many observations.
J.V. and Š.G. 
thank  the Slovak Grant Agency for Science VEGA 
(grant 1/0530/22).
TRAPPIST is funded by the Belgian Fund for Scientific Research (FNRS) under the grant PDR T.0120.21. E.J. is a FNRS Senior Research Associate
This research has made use of the Keck Observatory Archive (KOA),
which is operated by the W. M. Keck Observatory and the NASA
Exoplanet Science Institute (NExScI), under contract with the
National Aeronautics and Space Administration.
Some of the data presented herein were obtained at Keck Observatory, which is a private 501(c)3 non-profit organization operated as a scientific partnership among the California Institute of Technology, the University of California, and the National Aeronautics and Space Administration. The Observatory was made possible by the generous financial support of the W. M. Keck Foundation. 
The authors wish to recognize and acknowledge the very significant cultural role and reverence that the summit of Maunakea has always had within the Native Hawaiian community. We are most fortunate to have the opportunity to conduct observations from this mountain.
This publication makes use of data products from the Wide-field Infrared Survey Explorer, which is a joint project of the University of California, Los Angeles, and the Jet Propulsion Laboratory/California Institute of Technology, funded by the National Aeronautics and Space Administration.
This work is based in part on observations made with the Spitzer
Space Telescope, which was operated by the Jet Propulsion Laboratory, California
Institute of Technology under a contract with NASA.
The LBT is an international collaboration among institutions in the United States, Italy and Germany. LBT Corporation Members are: The University of Arizona on behalf of the Arizona Board of Regents; Istituto Nazionale di Astrofisica, Italy; LBT Beteiligungsgesellschaft, Germany, representing the Max-Planck Society, The Leibniz Institute for Astrophysics Potsdam, and Heidelberg University; The Ohio State University, representing OSU, University of Notre Dame, University of Minnesota and University of Virginia. This publication has made use of the Canadian Astronomy Data Centre moving object search tool \citep{CADCmovingobject}.
This research used the
\miriade \citep{2009-EPSC-Berthier},
\ssodnet \citep{2023A&A...671A.151B},
and
\topcat
\citep{2005ASPC..347...29T}
Virtual Observatory tools.
Some of the work presented here is based on observations collected
at the European Organisation for Astronomical Research in the Southern Hemisphere under ESO program 074.C-0052 and 089.C-0944 (PI
Marchis), 71.C-0669, 072.C-0753 (PI:Merline).
This research is based on observations made with the NASA/ESA Hubble Space Telescope obtained from the Space Telescope Science Institute, which is operated by the Association of Universities for Research in Astronomy, Inc., under NASA contract NAS 5–26555. These observations are associated with program(s) 10165 and 9747.



\ifx\destination\aanda
 \bibliographystyle{aux/aa} 
\fi


\bibliography{aux/bib, aux/morebib}

\clearpage

\appendix
\onecolumn
\section{Astrometry and photometry of the VWBA satellites\label{app:obs:sat}}

We list in the following tables the relative astrometry and
photometry of the satellites of
(379) Huenna (\Cref{tab:huennagenoid}),
(2577) Litva (\Cref{tab:litvagenoid}),
(3548) Eurybates(\Cref{tab:eurybatesgenoid},
(4674) Pauling (\Cref{tab:paulinggenoid}), 
(3749) Balam (\Cref{tab:balamgenoid}),
(17246) Christophedumas (\Cref{tab:positionschris}), and
(22899) Alconrad (\Cref{tab:positionsal}).

\begin{center}
  \begin{longtable}{cclllrrrrrrr}
  \caption[Astrometry of Huenna]{Astrometry of (379) Huenna's satellite S/2003 (379) 1.
    \label{tab:huennagenoid}
  }\\

    \hline\hline
     Date & UTC & Tel. & Cam. & Filter &
     \multicolumn{1}{c}{$X_o$} &
     \multicolumn{1}{c}{$Y_o$} &
     \multicolumn{1}{c}{$X_{o-c}$} &
     \multicolumn{1}{c}{$Y_{o-c}$} &
     \multicolumn{1}{c}{$\sigma$} &
     \multicolumn{1}{c}{$\Delta M$} &
     \multicolumn{1}{c}{$\delta M$} \\
    &&&&& 
     \multicolumn{1}{c}{(mas)} & \multicolumn{1}{c}{(mas)} & 
     \multicolumn{1}{c}{(mas)} & \multicolumn{1}{c}{(mas)} & 
     \multicolumn{1}{c}{(mas)} & 
     \multicolumn{1}{c}{(mag)} & \multicolumn{1}{c}{(mag)}  \\ 
    \hline
    \endfirsthead

    \multicolumn{11}{c}{{\tablename\ \thetable{} -- continued from previous page}} \\ 
    \hline\hline
     Date & UTC & Tel. & Cam. & Filter &
     \multicolumn{1}{c}{$X_o$} &
     \multicolumn{1}{c}{$Y_o$} &
     \multicolumn{1}{c}{$X_{o-c}$} &
     \multicolumn{1}{c}{$Y_{o-c}$} &
     \multicolumn{1}{c}{$\sigma$} &
     \multicolumn{1}{c}{$\Delta M$} &
     \multicolumn{1}{c}{$\delta M$} \\
    &&&&& 
     \multicolumn{1}{c}{(mas)} & \multicolumn{1}{c}{(mas)} & 
     \multicolumn{1}{c}{(mas)} & \multicolumn{1}{c}{(mas)} & 
     \multicolumn{1}{c}{(mas)} & 
     \multicolumn{1}{c}{(mag)} & \multicolumn{1}{c}{(mag)}  \\ 
    \hline
    \endhead

    \hline \multicolumn{11}{r}{{Continued on next page}} \\ \hline
    \endfoot

    \hline
    \endlastfoot
2003-08-14 & 13:04:25.90 & Keck II & NIRC2 & H & 501.6 & 737.0 & 0.3 & 4.0 & 10.0 & 8.1 & 0.6  \\ 
2003-08-14 & 14:07:52.88 & Keck II & NIRC2 & H & 506.3 & 737.3 & -3.3 & -0.3 & 10.0 & 7.8 & 0.3  \\ 
2003-08-15 & 13:34:46.63 & Keck II & NIRC2 & Kp & 693.1 & 838.7 & 0.4 & 0.6 & 10.0 & 7.0 & 0.1  \\ 
2003-08-15 & 13:38:59.65 & Keck II & NIRC2 & Kp & 696.3 & 841.7 & 3.1 & 3.3 & 10.0 & 7.1 & 0.1  \\ 
2003-08-17 & 14:53:55.04 & Keck II & NIRC2 & Kp & 1070.0 & 1028.6 & 1.2 & -7.3 & 10.0 & 7.1 & 0.1  \\ 
2003-08-18 & 14:19:56.05 & Keck II & NIRC2 & Kp & 1243.2 & 1116.1 & 1.6 & -6.7 & 10.0 & 6.9 & 0.0  \\ 
2004-12-08 & 07:00:25.58 & ESO-VLT & NACO & Ks & 1771.2 & 114.2 & -8.2 & -10.9 & 13.3 & 6.3 & 0.6  \\ 
2004-12-08 & 07:08:31.34 & ESO-VLT & NACO & Ks & 1783.4 & 130.1 & 4.1 & 4.8 & 13.3 & 7.4 & 0.8  \\ 
2004-12-09 & 06:29:18.80 & ESO-VLT & NACO & Ks & 1759.3 & 147.5 & 13.6 & -2.7 & 13.3 & 6.5 & 0.4  \\ 
2004-12-09 & 06:41:52.74 & ESO-VLT & NACO & Ks & 1743.5 & 146.8 & -1.8 & -3.7 & 13.3 & 6.9 & 0.0  \\ 
2004-12-10 & 06:45:20.88 & ESO-VLT & NACO & Ks & 1702.2 & 176.5 & 0.8 & 1.1 & 13.3 & 6.4 & 0.6  \\ 
2004-12-10 & 06:51:28.80 & ESO-VLT & NACO & Ks & 1697.7 & 174.4 & -3.5 & -1.1 & 13.3 & 6.7 & 0.0  \\ 
2004-12-14 & 05:28:42.92 & ESO-VLT & NACO & Ks & 1447.4 & 264.8 & -2.2 & 1.3 & 13.3 & 6.7 & 0.1  \\ 
2004-12-14 & 07:08:56.99 & ESO-VLT & NACO & Ks & 1448.6 & 266.8 & 4.4 & 1.9 & 13.3 & 6.7 & 0.1  \\ 
2004-12-15 & 05:20:24.80 & ESO-VLT & NACO & Ks & 1370.4 & 285.3 & 1.0 & 2.6 & 13.3 & 6.9 & 0.1  \\ 
2004-12-28 & 05:36:53.11 & ESO-VLT & NACO & Ks & -21.5 & 396.2 & 2.1 & 0.3 & 13.3 & 7.1 & 0.3  \\ 
2004-12-28 & 07:41:12.80 & ESO-VLT & NACO & Ks & -41.1 & 397.2 & -7.3 & 1.5 & 13.3 & 7.0 & 0.4  \\ 
2004-12-29 & 05:13:34.33 & ESO-VLT & NACO & Ks & -141.3 & 391.0 & -1.3 & -2.3 & 13.3 & 6.3 & 0.0  \\ 
2005-01-18 & 03:58:33.89 & ESO-VLT & NACO & Ks & -1924.7 & 65.5 & 3.9 & 5.1 & 13.3 & 7.0 & 0.1  \\ 
2005-01-18 & 06:17:33.16 & ESO-VLT & NACO & Ks & -1931.5 & 56.4 & 0.3 & -1.6 & 13.3 & 6.7 & 0.1  \\ 
2005-01-21 & 02:25:25.00 & ESO-VLT & NACO & Ks & -1991.4 & -15.5 & -4.6 & -2.4 & 13.3 & 6.8 & 0.5  \\ 
2005-01-25 & 04:51:31.42 & ESO-VLT & NACO & Ks & -1916.0 & -110.6 & -2.3 & -1.1 & 13.3 & 7.4 & 0.0  \\ 
2005-01-26 & 02:47:39.26 & ESO-VLT & NACO & Ks & -1868.8 & -127.0 & 0.7 & 1.8 & 13.3 & 6.9 & 0.0  \\ 
2005-01-26 & 05:10:41.18 & ESO-VLT & NACO & Ks & -1860.3 & -127.9 & 3.7 & 3.0 & 13.3 & 7.4 & 0.3  \\ 
2005-01-27 & 03:10:50.52 & ESO-VLT & NACO & Ks & -1799.4 & -146.9 & 7.9 & 2.0 & 13.3 & 7.1 & 0.2  \\ 
2005-01-28 & 03:04:35.77 & ESO-VLT & NACO & Ks & -1734.0 & -166.3 & -1.2 & 0.5 & 13.3 & 6.7 & 0.3  \\ 
2005-01-28 & 03:13:54.04 & ESO-VLT & NACO & H & -1735.4 & -159.6 & -3.1 & 7.4 & 13.3 & 7.4 & 0.1  \\ 
2005-02-02 & 03:09:16.21 & ESO-VLT & NACO & Ks & -1147.8 & -222.3 & 1.8 & 0.8 & 13.3 & 7.1 & 0.0  \\ 
2005-02-02 & 05:09:29.47 & ESO-VLT & NACO & Ks & -1134.4 & -225.9 & 2.6 & -2.4 & 13.3 & 7.0 & 0.1  \\ 
2005-02-04 & 02:41:04.73 & ESO-VLT & NACO & Ks & -823.5 & -223.4 & 6.1 & 1.9 & 13.3 & 7.0 & 0.2  \\ 
2005-02-04 & 04:05:56.80 & ESO-VLT & NACO & Ks & -819.9 & -224.2 & -0.4 & 1.0 & 13.3 & 6.8 & 0.1  \\ 
2005-02-16 & 01:21:08.59 & ESO-VLT & NACO & Ks & 1221.4 & -4.4 & -3.7 & -0.4 & 13.3 & 6.7 & 0.3  \\ 
2009-11-28 & 12:13:35.10 & Keck II & NIRC2 & Kp & -1672.8 & -616.3 & -4.8 & 6.6 & 10.0 & 7.2 & 0.1  \\ 
2009-11-28 & 12:51:36.23 & Keck II & NIRC2 & Kp & -1673.5 & -615.8 & -8.5 & 5.7 & 10.0 & 7.0 & 0.0  \\ 
2012-02-03 & 13:18:43.99 & Keck II & NIRC2 & Kp & 740.2 & -551.1 & -6.5 & -2.7 & 10.0 & 6.9 & 0.4  \\ 
2012-02-03 & 13:33:28.73 & Keck II & NIRC2 & Kp & 739.6 & -552.9 & -7.9 & -4.4 & 10.0 & 7.0 & 0.1  \\ 
2012-03-04 & 10:33:13.88 & Keck II & NIRC2 & Kp & -1225.8 & 894.4 & 4.1 & 17.5 & 10.0 & 7.4 & 0.3  \\ 
2012-03-04 & 10:46:09.13 & Keck II & NIRC2 & Kp & -1227.3 & 887.6 & 3.4 & 10.4 & 10.0 & 7.1 & 0.6  \\ 
2012-03-29 & 07:16:52.32 & Keck II & NIRC2 & Kp & -1881.0 & 656.1 & -0.5 & 2.7 & 10.0 & 6.9 & 0.0  \\ 
2012-06-02 & 00:51:58.54 & ESO-VLT & NACO & Ks & -1568.6 & 825.2 & -13.4 & -18.9 & 13.3 & 7.0 & 0.3  \\ 
2014-12-07 & 05:38:57.20 & Keck II & NIRC2 & Ks & -554.1 & -585.5 & -3.5 & -1.4 & 10.0 & 6.7 & 0.0  \\ 
2021-11-18 & 15:05:27.27 & Keck II & NIRC2 & H & -1859.6 & 175.3 & 4.4 & -12.7 & 10.0 & 8.2 & 0.1  \\ 
\end{longtable}
\footnotesize{Date, mid-observing time (UTC), telescope, camera, filter, 
    astrometry ($X$ is aligned with Right Ascension, and $Y$ with Declination, and 
    $o$ and $c$ indices stand for observed and computed positions), \KM{astrometric uncertainty in $X$ and $Y$ ($\sigma$) which are based on the pixel size and performance of the instrument,}
    and photometry (magnitude difference $\Delta M$ with uncertainty $\delta M$).}
\end{center}

\begin{center}
  \begin{longtable}{cclllrrrrrrr}
  \caption[Astrometry of Litva]{Astrometry of (2577) Litva's satellite S/2012 (2577) 1, \KM{see Table \ref{tab:huennagenoid} for a description of the columns}.
    \label{tab:litvagenoid}
  }\\

    \hline\hline
     Date & UTC & Tel. & Cam. & Filter &
     \multicolumn{1}{c}{$X_o$} &
     \multicolumn{1}{c}{$Y_o$} &
     \multicolumn{1}{c}{$X_{o-c}$} &
     \multicolumn{1}{c}{$Y_{o-c}$} &
     \multicolumn{1}{c}{$\sigma$} &
     \multicolumn{1}{c}{$\Delta M$} &
     \multicolumn{1}{c}{$\delta M$} \\
    &&&&& 
     \multicolumn{1}{c}{(mas)} & \multicolumn{1}{c}{(mas)} & 
     \multicolumn{1}{c}{(mas)} & \multicolumn{1}{c}{(mas)} & 
     \multicolumn{1}{c}{(mas)} & 
     \multicolumn{1}{c}{(mag)} & \multicolumn{1}{c}{(mag)}  \\ 
    \hline
    \endfirsthead

    \multicolumn{11}{c}{{\tablename\ \thetable{} -- continued from previous page}} \\ 
    \hline\hline
     Date & UTC & Tel. & Cam. & Filter &
     \multicolumn{1}{c}{$X_o$} &
     \multicolumn{1}{c}{$Y_o$} &
     \multicolumn{1}{c}{$X_{o-c}$} &
     \multicolumn{1}{c}{$Y_{o-c}$} &
     \multicolumn{1}{c}{$\sigma$} &
     \multicolumn{1}{c}{$\Delta M$} &
     \multicolumn{1}{c}{$\delta M$} \\
    &&&&& 
     \multicolumn{1}{c}{(mas)} & \multicolumn{1}{c}{(mas)} & 
     \multicolumn{1}{c}{(mas)} & \multicolumn{1}{c}{(mas)} & 
     \multicolumn{1}{c}{(mas)} & 
     \multicolumn{1}{c}{(mag)} & \multicolumn{1}{c}{(mag)}  \\ 
    \hline
    \endhead

    \hline \multicolumn{11}{r}{{Continued on next page}} \\ \hline
    \endfoot

    \hline
    \endlastfoot
2004-03-04 & 09:51:35.99 & ESO-VLT & NACO & Ks & 174.6 & -0.3 & 1.3 & 2.0 & 27.1 & 6.4 & 0.0  \\ 
2012-06-27 & 07:41:16.71 & Keck II & NIRC2 & Kp & 202.2 & -19.1 & -0.8 & 2.7 & 10.0 & 3.0 & 0.1  \\ 
2012-06-27 & 07:52:35.87 & Keck II & NIRC2 & H & 200.4 & -23.1 & -2.6 & -1.2 & 10.0 & 3.1 & 0.2  \\ 
2012-06-27 & 08:01:13.24 & Keck II & NIRC2 & J & 197.3 & -18.9 & -5.6 & 3.0 & 10.0 & 3.5 & 0.4  \\ 
2012-06-27 & 09:23:57.44 & Keck II & NIRC2 & Kp & 200.1 & -23.8 & -2.4 & -1.7 & 10.0 & 4.1 & 0.1  \\ 
2012-08-11 & 06:36:45.38 & Keck II & NIRC2 & Kp & -201.1 & -49.3 & 8.8 & -0.8 & 10.0 & 3.8 & 0.1  \\ 
2012-08-11 & 06:44:44.32 & Keck II & NIRC2 & Kp & -201.9 & -46.1 & 8.0 & 2.4 & 10.0 & 3.6 & 0.1  \\ 
2012-08-16 & 06:03:29.88 & Keck II & NIRC2 & Kp & -234.5 & -44.9 & -2.7 & -2.3 & 10.0 & 3.1 & 0.0  \\ 
2013-08-26 & 13:59:01.66 & Keck II & NIRC2 & Kp & -243.2 & 30.8 & -1.2 & 1.2 & 10.0 & 3.2 & 0.1  \\ 
2013-08-26 & 14:05:55.32 & Keck II & NIRC2 & H & -238.3 & 26.8 & 3.7 & -2.8 & 10.0 & 3.1 & 0.3  \\ 
2013-08-26 & 14:11:23.66 & Keck II & NIRC2 & Kp & -244.3 & 32.1 & -2.3 & 2.5 & 10.0 & 3.5 & 0.1  \\ 
2013-08-26 & 15:16:01.64 & Keck II & NIRC2 & Kp & -242.8 & 32.4 & -1.0 & 2.8 & 10.0 & 3.3 & 0.1  \\ 
2013-08-27 & 12:10:53.70 & Keck II & NIRC2 & Kp & -238.0 & 24.3 & 0.2 & -6.0 & 10.0 & 2.8 & 0.1  \\ 
2013-08-27 & 12:16:00.94 & Keck II & NIRC2 & H & -237.2 & 33.1 & 1.0 & 2.8 & 10.0 & 3.4 & 0.1  \\ 
2013-08-27 & 12:19:18.75 & Keck II & NIRC2 & H & -240.7 & 30.6 & -2.5 & 0.3 & 10.0 & 3.0 & 0.1  \\ 
2013-08-27 & 12:21:13.16 & Keck II & NIRC2 & H & -241.2 & 30.8 & -3.0 & 0.5 & 10.0 & 3.0 & 0.1  \\ 
2013-08-27 & 12:24:12.19 & Keck II & NIRC2 & Kp & -238.7 & 30.7 & -0.5 & 0.4 & 10.0 & 3.0 & 0.1  \\ 
2013-08-27 & 15:03:46.99 & Keck II & NIRC2 & Kp & -239.0 & 31.5 & -1.3 & 1.1 & 10.0 & 2.8 & 0.0  \\ 
2013-08-27 & 15:07:54.30 & Keck II & NIRC2 & H & -236.4 & 32.0 & 1.3 & 1.6 & 10.0 & 2.9 & 0.0  \\ 
2013-10-12 & 08:56:11.16 & LBTO & PISCES & H & 439.1 & -102.5 & 0.8 & -7.3 & 19.2 & 4.4 & 0.3  \\ 
2013-10-25 & 08:47:47.60 & Keck II & NIRC2 & H & 594.8 & -103.7 & 1.1 & 2.0 & 10.0 & 3.3 & 0.0  \\ 
2013-10-25 & 08:51:27.44 & Keck II & NIRC2 & H & 597.6 & -103.9 & 3.8 & 1.8 & 10.0 & 3.3 & 0.0  \\ 
  \end{longtable}
\end{center}

\begin{center}
  \begin{longtable}{cclllrrrrrrr}
  \caption[Astrometry of Queta]{Astrometry of Eurybates' satellite Queta, \KM{see Table \ref{tab:huennagenoid} for a description of the columns}.
    \label{tab:eurybatesgenoid}
  }\\

    \hline\hline
     Date & UTC & Tel. & Cam. & Filter &
     \multicolumn{1}{c}{$X_o$} &
     \multicolumn{1}{c}{$Y_o$} &
     \multicolumn{1}{c}{$X_{o-c}$} &
     \multicolumn{1}{c}{$Y_{o-c}$} &
     \multicolumn{1}{c}{$\sigma$} &
     \multicolumn{1}{c}{$\Delta M$} &
     \multicolumn{1}{c}{$\delta M$} \\
    &&&&& 
     \multicolumn{1}{c}{(mas)} & \multicolumn{1}{c}{(mas)} & 
     \multicolumn{1}{c}{(mas)} & \multicolumn{1}{c}{(mas)} & 
     \multicolumn{1}{c}{(mas)} & 
     \multicolumn{1}{c}{(mag)} & \multicolumn{1}{c}{(mag)}  \\ 
    \hline
    \endfirsthead

    \multicolumn{11}{c}{{\tablename\ \thetable{} -- continued from previous page}} \\ 
    \hline\hline
     Date & UTC & Tel. & Cam. & Filter &
     \multicolumn{1}{c}{$X_o$} &
     \multicolumn{1}{c}{$Y_o$} &
     \multicolumn{1}{c}{$X_{o-c}$} &
     \multicolumn{1}{c}{$Y_{o-c}$} &
     \multicolumn{1}{c}{$\sigma$} &
     \multicolumn{1}{c}{$\Delta M$} &
     \multicolumn{1}{c}{$\delta M$} \\
    &&&&& 
     \multicolumn{1}{c}{(mas)} & \multicolumn{1}{c}{(mas)} & 
     \multicolumn{1}{c}{(mas)} & \multicolumn{1}{c}{(mas)} & 
     \multicolumn{1}{c}{(mas)} & 
     \multicolumn{1}{c}{(mag)} & \multicolumn{1}{c}{(mag)}  \\ 
    \hline
    \endhead

    \hline \multicolumn{11}{r}{{Continued on next page}} \\ \hline
    \endfoot

    \hline
    \endlastfoot
2018-09-12 & 09:45:12.96 & HST & WFC3 & F555W & -484.9 & -103.0 * & 15.2 & -6.2 & 11.0 & 1.0 & 1.0  \\ 
2018-09-14 & 09:26:21.12 & HST & WFC3 & F555W & -430.1 & -67.0 * & 4.3 & -21.8 & 16.0 & 1.0 & 1.0  \\ 
2020-01-03 & 07:33:27.36 & HST & WFC3 & F350LP & -396.5 & -408.0 * & 1.0 & -5.2 & 5.0 & 1.0 & 1.0  \\ 
2020-07-19 & 20:40:16.32 & HST & WFC3 & F350LP & 430.5 & 460.0 & 6.6 & -0.2 & 3.0 & 1.0 & 1.0  \\ 
2020-10-12 & 19:08:15.36 & HST & WFC3 & F606W/F814W & 504.1 & 546.0 & -3.0 & -7.7 & 5.0 & 1.0 & 1.0  \\ 
2020-10-16 & 15:17:08.16 & HST & WFC3 & F606W/F814W & 438.2 & 487.0 & -5.3 & -4.4 & 15.0 & 1.0 & 1.0  \\ 
2020-11-24 * & 08:40:07.68 & HST & WFC3 & F606W/F814W & -551.4 & -605.0 & 1.3 & -5.9 & 5.0 & 1.0 & 1.0  \\ 
2021-02-10 & 06:33:59.04 & HST & WFC3 & F350LP & -440.3 & -462.0 & -4.6 & 4.1 & 5.0 & 1.0 & 1.0  \\ 
2021-02-12 & 15:45:12.96 & HST & WFC3 & F350LP & -434.1 & -476.0 & 9.3 & -3.3 & 4.0 & 1.0 & 1.0  \\ 
  \end{longtable}
  \footnotesize{Reproduced from \cite{2021-Brown-Eurybates} with minor corrections, \KM{astrometry with corrected sign errors is marked with a '*', as are corrected dates. Positions displayed in Figure 1 and JD in Table 2 of \cite{2021-Brown-Eurybates} are all correct}.}
\end{center}

\begin{center}
  \begin{longtable}{cclllrrrrrrr}
  \caption[Astrometry of (4674) Pauling]{Astrometry of (4674) Pauling's satellite S/2004 (4674) 1, \KM{see Table \ref{tab:huennagenoid} for a description of the columns}.
    \label{tab:paulinggenoid}
  }\\

    \hline\hline
     Date & UTC & Tel. & Cam. & Filter &
     \multicolumn{1}{c}{$X_o$} &
     \multicolumn{1}{c}{$Y_o$} &
     \multicolumn{1}{c}{$X_{o-c}$} &
     \multicolumn{1}{c}{$Y_{o-c}$} &
     \multicolumn{1}{c}{$\sigma$} &
     \multicolumn{1}{c}{$\Delta M$} &
     \multicolumn{1}{c}{$\delta M$} \\
     &&&&& 
     \multicolumn{1}{c}{(mas)} & \multicolumn{1}{c}{(mas)} & 
     \multicolumn{1}{c}{(mas)} & \multicolumn{1}{c}{(mas)} & 
     \multicolumn{1}{c}{(mas)} & 
     \multicolumn{1}{c}{(mag)} & \multicolumn{1}{c}{(mag)}  \\ 
    \hline
    \endfirsthead

    \multicolumn{11}{c}{{\tablename\ \thetable{} -- continued from previous page}} \\ 
    \hline\hline
     Date & UTC & Tel. & Cam. & Filter &
     \multicolumn{1}{c}{$X_o$} &
     \multicolumn{1}{c}{$Y_o$} &
     \multicolumn{1}{c}{$X_{o-c}$} &
     \multicolumn{1}{c}{$Y_{o-c}$} &
     \multicolumn{1}{c}{$\sigma$} &
     \multicolumn{1}{c}{$\Delta M$} &
     \multicolumn{1}{c}{$\delta M$} \\
     & & & & & 
     \multicolumn{1}{c}{(mas)} &
     \multicolumn{1}{c}{(mas)} & 
     \multicolumn{1}{c}{(mas)} &
     \multicolumn{1}{c}{(mas)} & 
     \multicolumn{1}{c}{(mas)} & 
     \multicolumn{1}{c}{(mag)} &
     \multicolumn{1}{c}{(mag)}  \\ 
    \hline
    \endhead

    \hline \multicolumn{11}{r}{{Continued on next page}} \\ \hline
    \endfoot

    \hline
    \endlastfoot
2004-03-04 & 05:30:40.08 & ESO-VLT & NACO & Ks & -302.8 & 249.1 & 0.2 & 5.4 & 27.1 & 3.6 & 0.2  \\ 
2004-03-04 & 05:41:18.31 & ESO-VLT & NACO & H & -311.4 & 248.3 & -8.3 & 4.6 & 13.3 & 4.3 & 0.2  \\ 
2004-03-04 & 05:53:20.04 & ESO-VLT & NACO & J & -307.9 & 240.3 & -4.7 & -3.3 & 13.3 & 4.8 & 0.7  \\ 
2004-03-04 & 07:11:14.62 & ESO-VLT & NACO & Ks & -309.0 & 242.8 & -5.1 & -0.7 & 27.1 & 3.4 & 0.3  \\ 
2005-09-10 & 11:42:01.13 & Keck II & NIRC2 & Kp & 222.7 & -55.3 & -1.1 & 14.0 & 10.0 & 2.9 & 0.0  \\ 
2007-05-28 & 10:15:45.14 & Keck II & NIRC2 & Kp & -404.0 & -73.0 & -7.2 & -11.7 & 10.0 & 5.4 & 0.6  \\ 
2012-06-22 & 07:59:36.29 & Keck II & NIRC2 & Kp & -345.4 & -70.9 & 5.1 & 6.3 & 10.0 & 2.7 & 0.0  \\ 
2012-06-27 & 08:57:47.33 & Keck II & NIRC2 & Kp & -380.3 & -58.1 & -0.7 & 4.5 & 10.0 & 3.0 & 0.0  \\ 
2012-06-27 & 09:05:35.66 & Keck II & NIRC2 & H & -380.4 & -58.2 & -0.7 & 4.4 & 10.0 & 3.1 & 0.0  \\ 
2012-06-27 & 09:13:56.33 & Keck II & NIRC2 & J & -381.2 & -59.5 & -1.5 & 3.0 & 10.0 & 3.4 & 0.1  \\ 
2012-06-27 & 10:35:11.14 & Keck II & NIRC2 & Kp & -377.9 & -57.4 & 2.0 & 4.9 & 10.0 & 3.1 & 0.0  \\ 
2012-07-15 & 07:36:06.96 & Keck II & NIRC2 & Kp & -323.0 & 9.4 & 1.7 & -1.0 & 10.0 & 2.9 & 0.0  \\ 
2012-07-15 & 07:39:22.23 & Keck II & NIRC2 & H & -323.2 & 9.1 & 1.5 & -1.3 & 10.0 & 3.0 & 0.0  \\ 
2012-07-15 & 09:32:42.36 & Keck II & NIRC2 & Kp & -320.0 & 9.2 & 4.0 & -1.5 & 10.0 & 3.0 & 0.0  \\ 
  \end{longtable}
\end{center}

\begin{center}
  \begin{longtable}{cclllrrrrrrr}
  \caption[Astrometry of S/2002 (3749) 1]{Astrometry of (3749) Balam's satellite S/2002 (3749) 1, \KM{see Table \ref{tab:huennagenoid} for a description of the columns}.
    \label{tab:balamgenoid}
  }\\

    \hline\hline
     Date & UTC & Tel. & Cam. & Filter &
     \multicolumn{1}{c}{$X_o$} &
     \multicolumn{1}{c}{$Y_o$} &
     \multicolumn{1}{c}{$X_{o-c}$} &
     \multicolumn{1}{c}{$Y_{o-c}$} &
     \multicolumn{1}{c}{$\sigma$} &
     \multicolumn{1}{c}{$\Delta M$} &
     \multicolumn{1}{c}{$\delta M$} \\
    &&&&& 
     \multicolumn{1}{c}{(mas)} & \multicolumn{1}{c}{(mas)} & 
     \multicolumn{1}{c}{(mas)} & \multicolumn{1}{c}{(mas)} & 
     \multicolumn{1}{c}{(mas)} & 
     \multicolumn{1}{c}{(mag)} & \multicolumn{1}{c}{(mag)}  \\ 
    \hline
    \endfirsthead

    \multicolumn{11}{c}{{\tablename\ \thetable{} -- continued from previous page}} \\ 
    \hline\hline
     Date & UTC & Tel. & Cam. & Filter &
     \multicolumn{1}{c}{$X_o$} &
     \multicolumn{1}{c}{$Y_o$} &
     \multicolumn{1}{c}{$X_{o-c}$} &
     \multicolumn{1}{c}{$Y_{o-c}$} &
     \multicolumn{1}{c}{$\sigma$} &
     \multicolumn{1}{c}{$\Delta M$} &
     \multicolumn{1}{c}{$\delta M$} \\
    &&&&& 
     \multicolumn{1}{c}{(mas)} & \multicolumn{1}{c}{(mas)} & 
     \multicolumn{1}{c}{(mas)} & \multicolumn{1}{c}{(mas)} & 
     \multicolumn{1}{c}{(mas)} & 
     \multicolumn{1}{c}{(mag)} & \multicolumn{1}{c}{(mag)}  \\ 
    \hline
    \endhead

    \hline \multicolumn{11}{r}{{Continued on next page}} \\ \hline
    \endfoot

    \hline
    \endlastfoot
2004-12-03 & 04:00:53.62 & ESO/VLT & NACO & Ks & -303.5 & 83.9 & 1.6 & -2.7 & 27.1 & 3.4 & 0.0  \\ 
2004-12-07 & 03:00:08.90 & ESO/VLT & NACO & Ks & -339.5 & 83.0 & 7.0 & 2.4 & 27.1 & 6.0 & 0.0  \\ 
2004-12-09 & 02:44:09.63 & ESO/VLT & NACO & Ks & -377.1 & 76.4 & -17.6 & 0.0 & 27.1 & 6.7 & 0.1  \\ 
2004-12-09 & 02:57:06.37 & ESO/VLT & NACO & Ks & -354.1 & 82.0 & 5.4 & 5.6 & 27.1 & 5.3 & 0.1  \\ 
2004-12-10 & 02:43:19.05 & ESO/VLT & NACO & Ks & -358.8 & 80.1 & 5.3 & 6.1 & 27.1 & 3.7 & 0.0  \\ 
2004-12-14 & 02:46:59.11 & ESO/VLT & NACO & Ks & -393.3 & 41.3 & -21.8 & -22.2 & 27.1 & 4.9 & 0.2  \\ 
2004-12-14 & 03:23:21.02 & ESO/VLT & NACO & Ks & -358.1 & 67.4 & 13.4 & 3.9 & 27.1 & 4.5 & 0.2  \\ 
2004-12-20 & 01:11:00.37 & ESO/VLT & NACO & Ks & -350.9 & 46.2 & 2.2 & -0.4 & 27.1 & 4.2 & 0.0  \\ 
2004-12-20 & 03:57:24.97 & ESO/VLT & NACO & Ks & -348.9 & 54.1 & 3.5 & 7.9 & 27.1 & 4.2 & 0.1  \\ 
  \end{longtable}
\end{center}

\begin{center}
  \begin{longtable}{cclllrrrrrrr}
  \caption[Astrometry of YYY]{Astrometry of (17246) Christophedumas's satellite S/2004 (17246) 1, \KM{see Table \ref{tab:huennagenoid} for a description of the columns}.
    \label{tab:positionschris}
  }\\

    \hline\hline
     Date & UTC & Tel. & Cam. & Filter &
     \multicolumn{1}{c}{$X_o$} &
     \multicolumn{1}{c}{$Y_o$} &
     \multicolumn{1}{c}{$X_{o-c}$} &
     \multicolumn{1}{c}{$Y_{o-c}$} &
     \multicolumn{1}{c}{$\sigma$} &
     \multicolumn{1}{c}{$\Delta M$} &
     \multicolumn{1}{c}{$\delta M$} \\
    &&&&& 
     \multicolumn{1}{c}{(mas)} & \multicolumn{1}{c}{(mas)} & 
     \multicolumn{1}{c}{(mas)} & \multicolumn{1}{c}{(mas)} & 
     \multicolumn{1}{c}{(mas)} & 
     \multicolumn{1}{c}{(mag)} & \multicolumn{1}{c}{(mag)}  \\ 
    \hline
    \endfirsthead

    \multicolumn{11}{c}{{\tablename\ \thetable{} -- continued from previous page}} \\ 
    \hline\hline
     Date & UTC & Tel. & Cam. & Filter &
     \multicolumn{1}{c}{$X_o$} &
     \multicolumn{1}{c}{$Y_o$} &
     \multicolumn{1}{c}{$X_{o-c}$} &
     \multicolumn{1}{c}{$Y_{o-c}$} &
     \multicolumn{1}{c}{$\sigma$} &
     \multicolumn{1}{c}{$\Delta M$} &
     \multicolumn{1}{c}{$\delta M$} \\
    &&&&& 
     \multicolumn{1}{c}{(mas)} & \multicolumn{1}{c}{(mas)} & 
     \multicolumn{1}{c}{(mas)} & \multicolumn{1}{c}{(mas)} & 
     \multicolumn{1}{c}{(mas)} & 
     \multicolumn{1}{c}{(mag)} & \multicolumn{1}{c}{(mag)}  \\ 
    \hline
    \endhead

    \hline \multicolumn{11}{r}{{Continued on next page}} \\ \hline
    \endfoot

    \hline
    \endlastfoot
2004-01-14 & 10:05:10.00 & HST & ACS & F606W & -181.4 & 18.3 & 0.3 & 0.4 & 3.0 & 2.0 & 0.0  \\ 
2005-05-18 & 01:24:23.00 & HST & ACS & F775W & 108.7 & 4.0 & -0.2 & 0.2 & 3.0 & 2.0 & 5.8  \\ 
2005-05-25 & 12:31:13.00 & HST & ACS & F775W & -12.6 & 68.0 & 0.3 & 0.1 & 3.0 & 3.6 & 0.1  \\ 
2005-06-25 & 03:57:43.00 & HST & ACS & F775W & -79.3 & -21.7 & 0.1 & 0.0 & 3.0 & 1.9 & 0.1  \\ 
  \end{longtable}
\end{center}

\begin{center}
  \begin{longtable}{cclllrrrrrrr}
  \caption[Astrometry of Alconrad]{Astrometry of (22899) Alconrad's satellite Juliekaibarreto, \KM{see Table \ref{tab:huennagenoid} for a description of the columns}.
    \label{tab:positionsal}
  }\\

    \hline\hline
     Date & UTC & Tel. & Cam. & Filter &
     \multicolumn{1}{c}{$X_o$} &
     \multicolumn{1}{c}{$Y_o$} &
     \multicolumn{1}{c}{$X_{o-c}$} &
     \multicolumn{1}{c}{$Y_{o-c}$} &
     \multicolumn{1}{c}{$\sigma$} &
     \multicolumn{1}{c}{$\Delta M$} &
     \multicolumn{1}{c}{$\delta M$} \\
    &&&&& 
     \multicolumn{1}{c}{(mas)} & \multicolumn{1}{c}{(mas)} & 
     \multicolumn{1}{c}{(mas)} & \multicolumn{1}{c}{(mas)} & 
     \multicolumn{1}{c}{(mas)} & 
     \multicolumn{1}{c}{(mag)} & \multicolumn{1}{c}{(mag)}  \\ 
    \hline
    \endfirsthead

    \multicolumn{11}{c}{{\tablename\ \thetable{} -- continued from previous page}} \\ 
    \hline\hline
     Date & UTC & Tel. & Cam. & Filter &
     \multicolumn{1}{c}{$X_o$} &
     \multicolumn{1}{c}{$Y_o$} &
     \multicolumn{1}{c}{$X_{o-c}$} &
     \multicolumn{1}{c}{$Y_{o-c}$} &
     \multicolumn{1}{c}{$\sigma$} &
     \multicolumn{1}{c}{$\Delta M$} &
     \multicolumn{1}{c}{$\delta M$} \\
    &&&&& 
     \multicolumn{1}{c}{(mas)} & \multicolumn{1}{c}{(mas)} & 
     \multicolumn{1}{c}{(mas)} & \multicolumn{1}{c}{(mas)} & 
     \multicolumn{1}{c}{(mas)} & 
     \multicolumn{1}{c}{(mag)} & \multicolumn{1}{c}{(mag)}  \\ 
    \hline
    \endhead

    \hline \multicolumn{11}{r}{{Continued on next page}} \\ \hline
    \endfoot

    \hline
    \endlastfoot
2003-07-26 & 03:45:47.00 & HST & ACS & F606W & -119.2 & -79.5 & 0.6 & 0.3 & 3.0 & 2.3 & 0.0  \\ 
2004-11-01 & 09:14:22.00 & HST & ACS & F775W & 67.1 & -55.5 & 1.4 & -3.2 & 3.0 & 2.8 & 0.1  \\ 
2004-11-13 & 23:38:58.00 & HST & ACS & F775W & -91.6 & 34.9 & -0.3 & 0.4 & 3.0 & 2.4 & 0.0  \\ 
2004-12-21 & 10:21:48.00 & HST & ACS & F775W & 124.5 & -61.5 & 0.2 & 0.1 & 3.0 & 2.6 & 0.0  \\ 
2004-12-27 & 11:53:24.00 & HST & ACS & F775W & 90.8 & -42.9 & -1.2 & 2.7 & 3.0 & 2.4 & 0.1  \\ 
  \end{longtable}
\end{center}

\section{Infrared photometry\label{app:obs:ir}}
Tables \ref{tab:spitzerchris} and \ref{tab:spitzeral} contain infrared photometry measurements from Spitzer observations of (17246) Christophedumas and (22899) Alconrad, respectively.


\begin{table}[h!]
\caption{Spitzer measurements for (17246) Christophedumas.}
    \label{tab:spitzerchris}
    \begin{tabular}{llllll}
    \hline \hline
    Telescope & Observation Mid-Time & r (au)  &  $\Delta$ (au) &  $\alpha$(deg)\\
    \hline\
    Spitzer   & 2005-Aug-22 16:38    & 2.7815  & 2.4091      &  -21.2 \\
    \hline
               & Instrument/Band & $\lambda$ ($\mu$m) & flux density (mJy) &
    Meas. unc. (mJy) & Total unc. (mJy)\\
               & IRAC1 & 0.0410 & 0.0016 & 0.0020  \\
               & IRAC2 & 0.0471 & 0.0019 & 0.0024  \\
               & IRAC3  &  5.73 & 0.213 & 0.008 &  0.010\\
               & IRAC4 &  7.87  & 1.810 & 0.012  &   0.056\\
    \hline
    WISE      & 2010-Mar-14 16:30     & 2.7835  & 2.5899     &   +20.9 \\
    \hline
               & Instrument/Band & $\lambda$ ($\mu$m) & flux density (mJy) &
    Meas. unc. (mJy) & Total unc. (mJy)\\
               & W3  & 11.1  & 6.2   & -   & 0.6\\
               & W4 & 22.6    & 17.8  & -  & 4.1\\
    \hline
\end{tabular}
\end{table}


\begin{table}[h!]
\caption{Spitzer measurements for (22899) Alconrad.}
\label{tab:spitzeral}
    \begin{tabular}{llllll}
        \hline \hline
        Telescope & Observation Mid-Time & r (au)  &  $\Delta$ (au) &  $\alpha$(deg)\\
        \hline\
        Spitzer   & 2005-Dec-27 07:35    & 3.0524 &  2.8078 & +19.3 \\
        \hline
                   & Instrument/Band & $\lambda$ ($\mu$m) & flux density (mJy) &
        Meas. unc. (mJy) & Total unc. (mJy)\\
                   & IRAC1 & 0.0255 & 0.0005 & 0.0009  \\
                   & IRAC2 & 0.0252 & 0.0007 & 0.0010  \\
                   & IRAC3  & 0.0943 & 0.0037 & 0.0047 & 0.005 \\
                   & IRAC4 & 0.9275 & 0.0055  & 0.0284 & 0.028 \\
        \hline
        WISE      & 2010-Mar-11 02:00   &  2.9137 & 2.6496     &  +19.8\\
     
        \hline
                   & Instrument/Band & $\lambda$ ($\mu$m) & flux density (mJy) &
        Meas. unc. (mJy) & Total unc. (mJy)\\
                   & W3  & 11.1  & 6.09   & -   & 0.58\\
                   & W4 & 22.6    & 16.96  & -  & 1.14\\
        \hline
    \end{tabular}
\end{table}

\twocolumn
\clearpage
\section{NEATM fits \label{app:NEATM}}

\begin{figure}[h!]
    \centering
    \includegraphics[width=\columnwidth]{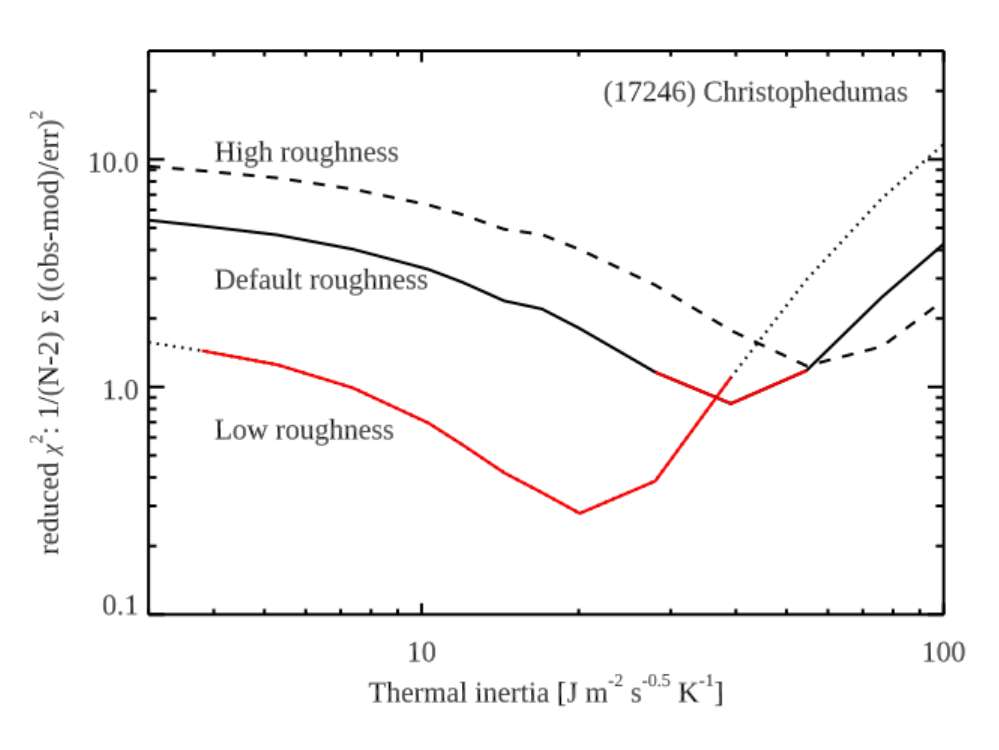}
    \caption{
    $\chi^2$ curves from fitting TPM predictions to the
WISE and Spitzer-IRAC data \KM{for asteroid (17246) Christophedumas}. Three different TPM solutions are shown: assuming
a low, (dotted line), intermediate (default value; see \cite{1998A&A...338..340M}), and high level (dashed line) of surface roughness. The red parts show acceptable TPM settings with possible thermal inertias in the range between 4 and 50 J m$^{-2}$ s$^{-0.5}$ K$^{-1}$.}
    \label{fig:CD_NEATM}
\end{figure}

\begin{figure}[h!]
    \centering
    \includegraphics[width=\columnwidth]{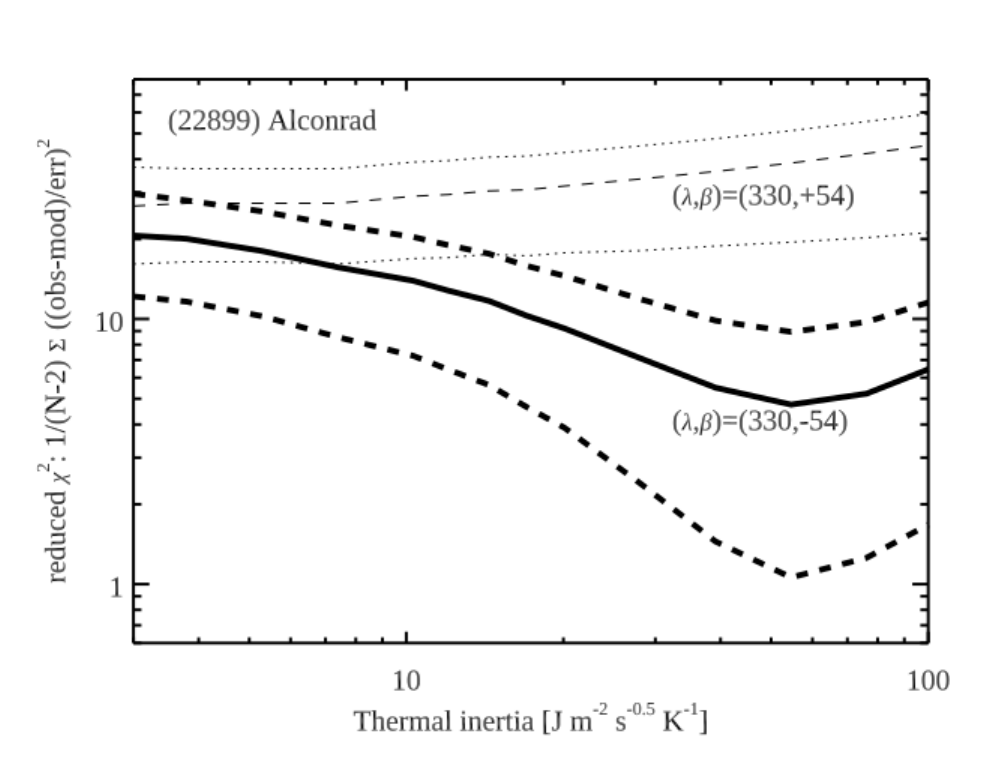}
    \caption{
    $\chi^2$ curves from fitting TPM predictions to the
WISE and Spitzer-IRAC data \KM{for asteroid (22899) Alconrad}. Three different TPM solutions are shown for a prograde rotation of the primary (top 3 thin lines), and another three lines
(thick dashed and solid lines) for the retrograde rotation case. Based on our radiometric study, the retrograde rotation is strongly favoured, possibly with a thermal inertia in the 30-100 J m$^{-2}$ s$^{-0.5}$ K$^{-1}$ range. The retrograde rotation is in agreement with the retrograde orbit of the satellite.}
    \label{fig:AC_NEATM}
\end{figure}

\end{document}